\documentclass[5p,times]{elsarticle}

\usepackage{amsmath,amsfonts}
\usepackage{array}
\usepackage[caption=false,font=normalsize,labelfont=sf,textfont=sf]{subfig}
\usepackage{textcomp}
\usepackage{stfloats}
\usepackage{url}
\usepackage{verbatim}
\usepackage{graphicx}
\usepackage{nicefrac}
\usepackage{tabularx}

\usepackage{siunitx}
\DeclareMathOperator*{\minimize}{minimize}

\newcommand{\R}{\mathbb{R}}

\newcommand{\setT}{\Gamma}
\newcommand{\setP}{\mathcal{P}}
\newcommand{\setR}{\mathcal{R}}
\newcommand{\guess}{\text{guess}}

\usepackage{xcolor}
\colorlet{punct}{red!60!black}
\definecolor{background}{HTML}{EEEEEE}
\definecolor{delim}{RGB}{20,105,176}
\colorlet{numb}{magenta!60!black}

\usepackage{listings}
\lstdefinelanguage{Json}{
	basicstyle=\footnotesize\ttfamily,
	numbers=none,
	numberstyle=\scriptsize,
	abovecaptionskip=\smallskipamount,
	belowcaptionskip=\smallskipamount,
	captionpos=b,
	stepnumber=1,
	numbersep=8pt,
	showstringspaces=false,
	breaklines=true,
	frame=lines,
	backgroundcolor=\color{background},
	literate=
	*{0}{{{\color{numb}0}}}{1}
	{1}{{{\color{numb}1}}}{1}
	{2}{{{\color{numb}2}}}{1}
	{3}{{{\color{numb}3}}}{1}
	{4}{{{\color{numb}4}}}{1}
	{5}{{{\color{numb}5}}}{1}
	{6}{{{\color{numb}6}}}{1}
	{7}{{{\color{numb}7}}}{1}
	{8}{{{\color{numb}8}}}{1}
	{9}{{{\color{numb}9}}}{1}
	{:}{{{\color{punct}{:}}}}{1}
	{,}{{{\color{punct}{,}}}}{1}
	{\{}{{{\color{delim}{\{}}}}{1}
	{\}}{{{\color{delim}{\}}}}}{1}
	{[}{{{\color{delim}{[}}}}{1}
	{]}{{{\color{delim}{]}}}}{1},
}
\lstdefinelanguage{Python}{
	basicstyle=\footnotesize\ttfamily,
	numbers=none,
	numberstyle=\scriptsize,
	abovecaptionskip=\smallskipamount,
	belowcaptionskip=\smallskipamount,
	captionpos=b,
	stepnumber=1,
	numbersep=8pt,
	showstringspaces=false,
	breaklines=true,
	frame=lines,
	backgroundcolor=\color{background},
	literate=
	*{0}{{{\color{numb}0}}}{1}
	{1}{{{\color{numb}1}}}{1}
	{2}{{{\color{numb}2}}}{1}
	{3}{{{\color{numb}3}}}{1}
	{4}{{{\color{numb}4}}}{1}
	{5}{{{\color{numb}5}}}{1}
	{6}{{{\color{numb}6}}}{1}
	{7}{{{\color{numb}7}}}{1}
	{8}{{{\color{numb}8}}}{1}
	{9}{{{\color{numb}9}}}{1}
	{:}{{{\color{punct}{:}}}}{1}
	{,}{{{\color{punct}{,}}}}{1}
	{\{}{{{\color{delim}{\{}}}}{1}
	{\}}{{{\color{delim}{\}}}}}{1}
	{[}{{{\color{delim}{[}}}}{1}
	{]}{{{\color{delim}{]}}}}{1},
}

\journal{Future Generation Computer Systems}

\begin{document}

\begin{frontmatter}

\title{Relativistic Digital Twin: Bringing the IoT to the Future}

\author[label1,label2]{Luca Sciullo\corref{cor1}}
\ead{luca.sciullo@unibo.it}
\author[label3]{Alberto De~Marchi}
\ead{alberto.demarchi@unibw.de}
\author[label1]{Angelo Trotta}
\ead{angelo.trotta5@unibo.it}
\author[label1,label2]{Federico Montori}
\ead{federico.montori2@unibo.it}
\author[label1,label2]{Luciano Bononi}
\ead{luciano.bononi@unibo.it}
\author[label1,label2]{Marco Di~Felice}
\ead{marco.difelice3@unibo.it}

\affiliation[label1]{%
    organization={Department of Computer Science and Engineering, University of Bologna},
    country={Italy},
}
\affiliation[label2]{%
    organization={Advanced Research Center on Electronic Systems, University of Bologna},
    country={Italy},
}
\affiliation[label3]{%
    organization={Institute of Applied Mathematics and Scientific Computing, Department of Aerospace Engineering, University of the Bundeswehr Munich},
    country={Germany},
}

\cortext[cor1]{Corresponding author}

\begin{abstract}
    Complex IoT ecosystems often require the usage of Digital Twins (DTs) of their physical assets in order to perform predictive analytics and simulate what-if scenarios. DTs are able to replicate IoT devices and adapt over time to their behavioral changes. However, DTs in IoT are typically tailored to a specific use case, without the possibility to seamlessly adapt to different scenarios. Further, the fragmentation of IoT poses additional challenges on how to deploy DTs in heterogeneous scenarios characterized by the usage of multiple data formats and IoT network protocols. In this paper, we propose the Relativistic Digital Twin (RDT) framework, through which we automatically generate general-purpose DTs of IoT entities and tune their behavioral models over time by constantly observing their real counterparts. The framework relies on the object representation via the Web of Things (WoT), to offer a standardized interface to each of the IoT devices as well as to their DTs. To this purpose, we extended the W3C WoT standard in order to encompass the concept of behavioral model and define it in the Thing Description (TD) through a new vocabulary. Finally, we evaluated the RDT framework over two disjoint use cases to assess its correctness and learning performance, i.e., the DT of a simulated smart home scenario with the capability of forecasting the indoor temperature, and the DT of a real-world drone with the capability of forecasting its trajectory in an outdoor scenario. Experiments show that the generated DT can estimate the behavior of its real counterpart after an observation stage, regardless of the considered scenario.
\end{abstract}


\end{frontmatter}

\section{Introduction}
The near future of the Internet of Things (IoT) paradigm is intertwined with the novel concept of Digital Twins (DTs). DTs are digital replicas of physical assets that evolve over time in order to mimic their behavior, their reaction to events and their internal structure as closely as possible through computational models \cite{AIAA2020digital}. In previous years, other entities, like simulators, emulators, digital shadows, mirror systems, or avatars, assumed this role. DTs are different from them in that (\textit{i}) they constantly receive inputs from the mirrored appliance(s) to adapt to its behavioral changes, as well as (\textit{ii}) they feed back their outputs into the real environment to provide guidance \cite{rasheed2020digital}.
The capability of DTs to reproduce real devices and entities makes them extremely important for a variety of IoT scenarios that take advantage of predictive and what-if analytics. In fact, DTs find their natural applicability in manufacturing and Industry 4.0 scenarios, where putting the physical asset at risk results in a massive loss in terms of resources \cite{kritzinger2018digital}.
Because of their crucial role, DTs are, most of the time, tailored to the appliance that they reproduce. This is obviously a safe choice, however, it results in additional engineering cost and effort when it is necessary to adapt the computational model to a new use case or a new asset \cite{niederer2021scaling}. As a matter of fact, nowadays there is a profound lack of general-purpose solutions that can offer adaptable DTs or automatically generate computational models able to synthesize heterogeneous scenarios. It is clear that a zero-knowledge solution is nearly impossible to obtain, but we can envision an automatic process that is able to fit a variety of scenarios given very little initial information. 
Moreover, the age-old problem of IoT fragmentation still persists in most of the IoT ecosystems and exacerbates the difficulties of introducing such automation for DTs, as devices may use different technologies and protocols to expose their data and services in a highly dynamic environment \cite{amelie2018}.
To overcome this issue, several solutions have been proposed over the last decade; one of the most supported is the Web of Things (WoT), recently standardized by the World Wide Web Consortium (W3C) \cite{W3Cwot}. The WoT abstracts each IoT device into a Web Thing (WT), a digital entity that exposes its functionalities (defined as \textit{affordances} in the official documentation \cite{W3Cwot}) via standardized Web technologies and semantics that vaguely resemble the REST concepts.
WTs are described via a Thing Description (TD), a machine-readable JSON-LD document that can be shared with other WTs in order to automate interactions in a WoT ecosystem. A number of Industrial IoT frameworks are now dedicated to embracing the concept of DTs, at the same time leveraging integration with the WoT. An example is Eclipse Ditto \cite{jackle2022ditto} or Eclipse Arrowhead \cite{varga2017making}, which presents a one-to-one bi-directional translation between service records and WTs \cite{zyrianoff2021two}.
These solutions exploit the WoT for abstracting heterogeneous DTs of physical assets and integrating them within a single application ecosystem, however, they do not take into account a homogenized representation of their behavioral models, leaving this aspect to the third parties.

In this paper we propose a solution to the two above-mentioned shortcomings of the current landscape of DTs in IoT: either (\textit{i}) DTs are too specialized for the use case they operate upon, or (\textit{ii}) interoperability-oriented IoT frameworks do not have the necessary expressiveness to encompass their model description. 
Specifically, we propose a framework that is based on the WoT, where all entities are abstracted into a single WT, and all interactions between entities are encoded as affordances. However, as an extension to the W3C WoT standard and as the main contribution of our work, each of the TDs is enriched with a parametric behavioral model that characterizes the dynamic of the target IoT appliance over time. Based on it, our framework is able to generate a DT for each of the represented WTs that presents the exact same affordances; in addition, by periodically monitoring the behavior of the real appliance, the framework fits the model parameters to mimic the physical asset over time. Eventually, the generated DTs will be indistinguishable from the real assets, as they both expose the exact same interface to other actors in the WoT ecosystem. 
A fully operational scenario aims to advocate a paradigm shift, where virtual entities can be queried -- and acted upon -- with a predictive connotation, by running them at future instants in time, as the DTs will be able to foretell the behavior of the asset. 
This vision is inspired by the ``Twin Paradox'' \cite{pesic2003einstein} in which one of the twins is sent forward in time and is eventually older than the other twin; in the same way here, the DT can interact with the other elements in the system without having to stick to the current moment.
This enables the construction of what-if scenarios, where we can spawn multiple DTs of the same WT, each of them with its own timeline. Each DT is then acted upon predictively, allowing us to ``bring the IoT to the future''.
Despite the shared birth between the DT and its physical counterpart, and because of the independent timeline of virtual entities, we name our framework the Relativistic Digital Twin (RDT) framework.

More precisely, this paper makes the following three contributions:
\begin{itemize}
	\item We propose an extension to the WoT TD syntax, in order to introduce the concept of behavioral model as well as its interactions; because certain entities in an ecosystem may influence each other, we also take into account that certain behavioral models may be dependent on the properties of other IoT objects in the same environment.
	\item We propose the RDT framework, which is able to generate DTs of each WT that includes a behavioral model in the ecosystem, by learning over time the model parameters through constant observation of the WT properties. At the end of the learning phase, the RDT framework is able to generate the Digital Twin Web Thing (DTWT), i.e.,  new WTs which are functionally equivalent to the real-world ones but are attached to the behavioral model rather than to the IoT physical asset. 
	\item We evaluate our proposal over two conceptually different use cases: a DT of a simulated scenario including the representation of a room with adjustable temperature and multiple sensors/actuators, and a DT of a real quadcopter drone able to reproduce its kinematic.
\end{itemize}

The rest of the paper is organized as follows: 
Section~\ref{sec:related} reviews the state-of-the-art concerning DT frameworks and the WoT, 
Section~\ref{sec:scenario} introduces the system model used in this work along with the scenario description and assumptions declaration,
Section~\ref{sec:model} outlines in detail how we extend the TD syntax to include behavioral models and their interactions,
Section~\ref{sec:architecture} illustrates the framework architecture and how we implemented all the components,
Section~\ref{sec:learning} describes the learning module and how each behavioral model is fitted to the real observations,
Section~\ref{sec:validation} introduces the two use cases on which we tested the framework and their results, and, finally,
Section~\ref{sec:conclusion} concludes the paper, outlining future works.

\section{Related Work} \label{sec:related}
\subsection{Digital Twin and Learning}
\label{sub:ml}

Adapting to behavioral changes is a prerogative of DTs, however, in a variety of scenarios, this is subject to multi-objective optimization problems that are mathematically intractable and for which only learning-based approaches and heuristics are viable solutions \cite{huang2021survey}. In some of these cases, the behavior of a physical asset is discrete and can be modeled as a state machine. When this happens, most solutions in the literature opt for Reinforcement Learning (RL) techniques applied over Markov Decision Processes (MDP), so that global KPIs can be autonomously reached by tuning model parameters over time. Examples can be found in \cite{wang2020adaptive} and \cite{jaensch2018reinforcement}, however, these works are solely designed for a single use case and cannot generalize. On the other hand, though, we note that the need for a generalized technique for generating DTs is extremely concrete and seen as a fundamental building block of the path ahead in this field \cite{minerva2020digital}.
In a first proposal in such a direction, we modeled the state space of discrete DTs as MDPs and provided a generic ``generator'' of DTs via the WoT paradigm
\cite{sciullo2022wotwins}. We used a combination of frequency-based past observation and deep neural networks to model the physical asset's behavior even in the presence of unseen states. Similarly, the work in \cite{kapteyn2021probabilistic} is a recent proposal that aims to move from one-off implementation of DTs to a generative paradigm, where formal models represent mathematically the physical object and learn its behavior according to its state space and what is being observed in the real world by its sensors. In detail, they propose a general framework based on a graphical model that estimates the state variables probabilistically according to partially observable Markov Chains. They use then the UAV context to evaluate the framework in the real world. This work however focuses on the mathematical models and gives little space to data representation, which is instead a core part of the present work.
In the same vein, we can find other proposals such as the work in \cite{agrell2021optimal}, based on Probabilistic Digital Twins (PDTs) that are built through Bayesian probabilistic frameworks. In this case, the key concept is that DTs must rely not only on \textit{observations}  coming from sensing experience but also on \textit{assumptions}, as in fact it is never possible to observe every single property that affects the physical asset, therefore, some of the observations must be replaced by mathematical models in order to capture uncertainties and to provide a minimum starting knowledge to the framework. In that work, such uncertainties are modeled as epistemic and aleatory within the proposed framework, which uses deep reinforcement learning for making predictions. Similarly, in this work, DTs are modeled through both assumptions (which define the base models for properties) and observations (which contribute to parameter tuning). However, we detach from state-based discrete properties, rather we take into account continuous properties that can be synthesized via regressive models. This applied to generalized DTs has not been tackled yet in literature, furthermore, we provide an accurate vocabulary for DTs representation via the WoT, introduced in the following section.

\subsection{Web of Things}
\label{sub:wot}

The world of IoT is known to be affected by technology and standard fragmentation, due to the rise of newer and newer technologies, to which current standards often cannot keep up the pace. This causes interoperability problems that need to be tackled at design time, at the same time tolerating the coexistence of heterogeneous systems. 
The WoT is a paradigm that proposes to extend IoT entities by using standard Web technologies \cite{sciullo2022survey}, which are solid and coherent throughout history. In this paper we will make use of the W3C WoT standard \cite{kovatsch2020web}, however, we will only use the term WoT to indicate it. In detail, the WoT standard abstracts a physical asset into a virtual counterpart called Web Thing (WT), which acts as a proxy between the Web world and the physical one. A WT exposes the capabilities and properties of the physical asset just like a Web Server, by providing a set of endpoints that can be queried, furthermore, it can perform actions actively against other WTs. For this reason, the software component embedding the WT is called ``servient''. The metadata describing all the interactions that a WT can handle is stored within a Thing Description (TD), a JSON-LD document that acts as the manifest of the WT by expressing its \textit{affordances}. These are:
\begin{itemize}
    \item \textit{Properties}, which correspond to state variables that can be fetched and changed at runtime.
    \item \textit{Actions}, which denote function calls to be performed by the WT, typically triggering a defined behavior.
    \item \textit{Events}, which are asynchronous and cause the WT to react to their occurrence.
\end{itemize}

By nature, the WoT paradigm provides an environment that fits very well with the implementation of DTs, which can easily be embedded into existing servients. However, besides our past work that we mentioned \cite{sciullo2022wotwins}, these concepts have rarely been paired in literature. An example is \cite{aguzzi2021modron}, where a ``Digital Twin WT'' is envisioned as a virtual entity derived from the aggregation of multiple physical ones. Differently, in \cite{privat2019wot}, the WoT is seen as the scaffolding to better grasp the intertwining of single components within structurally complex Systems-of-Systems, whilst in \cite{muralidharan2020designing} DTs are WTs that are automatically generated and managed within a container middleware, however, authors do not specify whether the behavior is also reproduced, as the focus is semantic interactions. 
DT is an extremely novel concept and very few research efforts have been directed toward automatic generation of DTs in an IoT generic context. Out of them, none implies or proposes a standard like the WoT to describe how to interact with them.

\section{System Model} \label{sec:scenario}
\begin{figure*}
	\centering
	\includegraphics[width=0.92\textwidth]{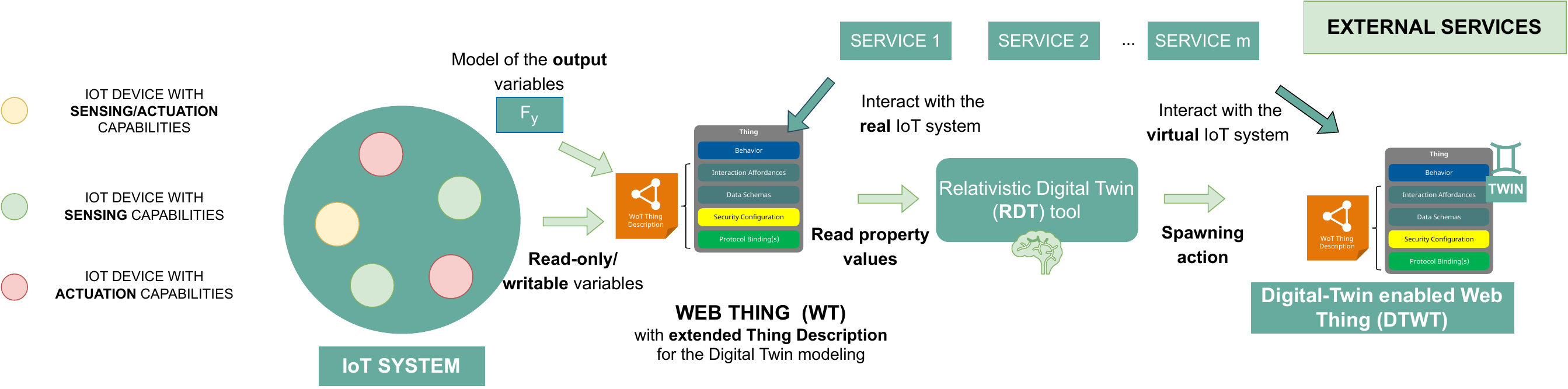}
	\caption{An IoT/WoT generic scenario and its DTWT concept.}
	\label{fig:scenario}
\end{figure*}

\begin{table}[tb]
	\centering
	\caption{\label{acronyms-table}Glossary of acronyms used in the paper.}
	\begin{tabularx}{\columnwidth}{c|X}
		Acronym & Meaning \\ \hline
		WoT     & W3C Web of Things: it refers to the set of standards for the Web of Things               \\
		WT      & Web Thing: a single Thing that was mapped with the W3C WoT                                   \\
		DT      & Digital Twin                                                                                 \\
		TD      & Thing Description: semantically-enriched description of a WT according to the WoT \\
		DTWT    & Digital Twin Web Thing: DT of a WT \\
		RDT     & Relativistic Digital Twin: the framework proposed in the paper
	\end{tabularx}
\end{table}

We now give a formal presentation of the generic setting for our framework, before moving to a more concrete and illustrative setting.
Let us consider a generic IoT scenario depicted in Figure~\ref{fig:scenario} and composed of $N \geq 1$ devices, $D = \{d_1, \dots, d_N\}$, with sensing and/or actuation capabilities. Each IoT device $d_i$ is associated to a list $S_i = \{s_{i,1}, s_{i,2}, \dots\}$ of time-dependent variables determining its state over time. Let $s_{i,j}(t)$ be the value of variable $s_{i,j}$ of device $d_i$ at time $t$, where the time $t \in \R$ is continuous. 
More specifically, we assume that for each device $d_i$, $S_i = R_i \bigcup W_i$, where:
\begin{itemize}
	\item $R_i = \{r_{i,1}, r_{i,2}, \dots\}$ indicates the set of \textit{read-only} variables. An instance can be the variable storing the measurement produced by a sensor unit, e.g., the temperature value. 
	\item $W_i = \{w_{i,1}, w_{i,2}, \dots\}$ indicates the set of \textit{writable} variables. An instance can be a tunable variable triggering an action: e.g., in the case of a led actuator $d_1$ having a single variable $w_{1,1}$ that defines the led luminescence, setting $w_{1,1}(t)$ equal to \texttt{\{True | False}\} corresponds to the action at time $t$ of turning \textit{on} or \textit{off} the light, respectively. Another instance can be represented by a configuration parameter of the device, e.g., the sampling frequency of a sensor unit.
\end{itemize}
This distinction allows to capture the difference between properties whose value directly depends on external actions, or is the result of some dynamics or relationships within the system of interest. While writable variables are essentially inputs applied to the system, read-only variables can be merely monitored, if observable at all. Note that this categorization is not unique and may arise from, and then reflect, the intended purpose of the DT.
Moreover, we consider that all IoT devices in our environment can have causal relationships with each other, i.e., the change of a writable variable for one device may affect the values of the  variables of other devices.
Given such premises, the overall goal of our study is how to support the automatic deployment of a DT of the whole IoT scenario, considered  as a single entity.
To this purpose, let  $R=\bigcup_i R_i$ be the set of the read-only system variables, $W=\bigcup_i W_i$ be the set of the writable system variables, and $S = R \bigcup W$. Moreover, let $R(t)$, $W(t)$, and $S(t)$ be the value at time $t$ for all the state variables in $R$, $W$, and $S$, respectively.
We call $Z = \{ z_1, z_2, \dots \}$, with $Z \subseteq R$, the set of \textit{output} system variables that the DT is in charge of prediction.
Our goal is to automatize the estimation of the (partially unreadable) system state based on previous actions and measurements.

\begin{figure}[t]
	\centering
	\includegraphics[width=0.95\linewidth]{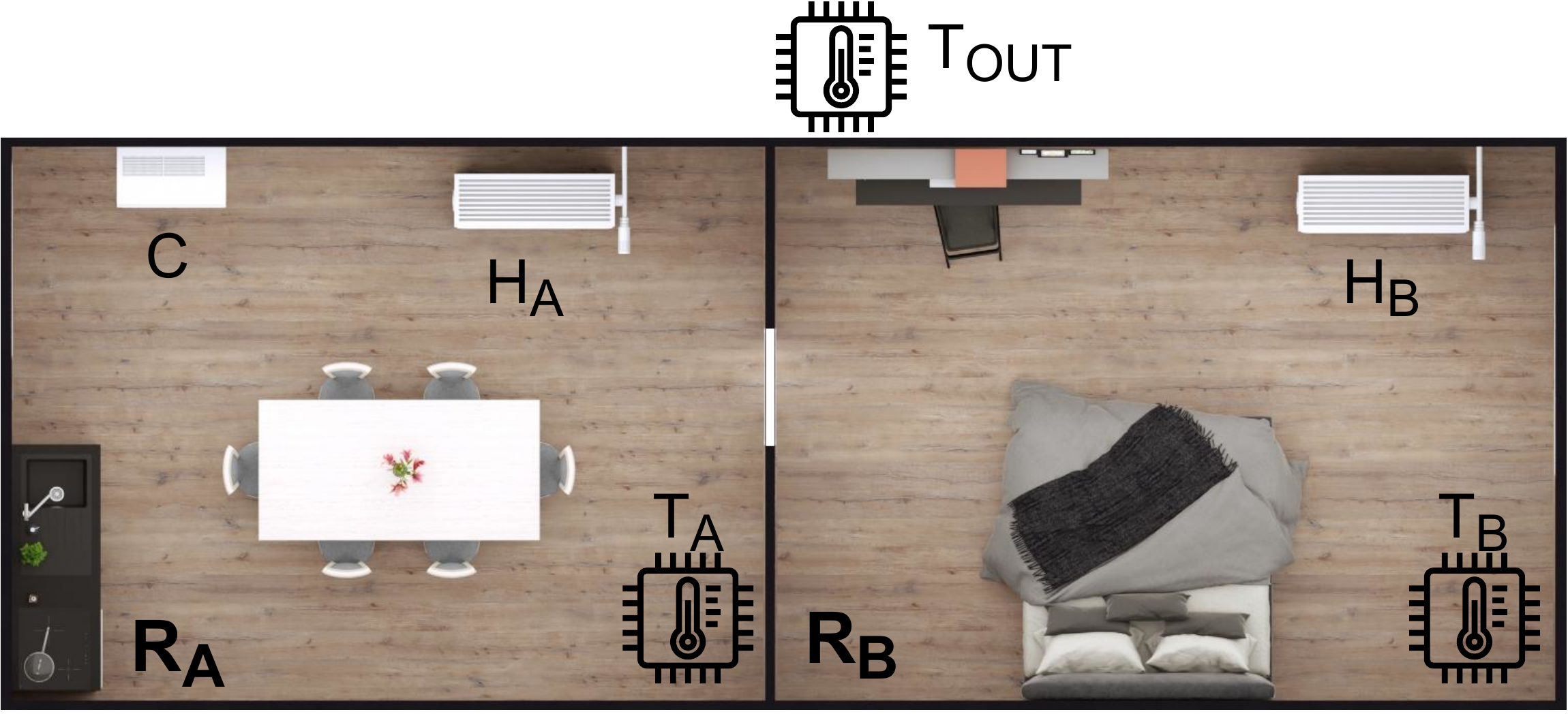}
	\caption{The room simulated scenario.}
	\label{fig:roomscenario}
\end{figure}

{It is worth mentioning that this model formalization is independent of a specific use case, and hence it can be easily applied to any IoT scenario that involves devices that interact with each other and that can be described through \textit{read-only}, \textit{writable} variables. This can be considered the only requirement for the target IoT System, since these variables must be mapped to affordances and models in a TD (as better described in Section\mbox{~\ref{sec:model}}) and enhanced with specific domain properties, that can be expressed through a semantic description. This is better depicted in the left part of Figure\mbox{~\ref{fig:scenario}}, where the overall methodology is presented: a generic IoT system composed of heterogeneous IoT devices with sensing and/or actuation capabilities is represented through a TD that is composed of some \textit{read-only} and \textit{writable} variables and the set of models for the \textit{output} variables and of the devices, encoded as affordances in the TD. This means that -for example- a \textit{smart vehicle} with its engine sensors, a \textit{smart room} with domotics devices, or a \textit{smart field} with its irrigation actuators share the same methodology of being mapped to a TD; the only difference is represented by the domain knowledge that enriches the TD affordances and that can be manually added to semantically describe the IoT System}.
To give concreteness to all these concepts, we considered the scenario depicted in Figure~\ref{fig:roomscenario}, developed in \mbox{\ref{sec:appendix}} and then evaluated in Section~\ref{sec:validation}. The IoT system in the exam is a closed smart home environment composed of two rooms, $R_A$ and $R_B$.  The full system is mapped to a single WT, exposing as properties the variables of the sensors/actuators.
In our case, each room is hosting a temperature sensor, $T_A$ and $T_B$, respectively. Each of the temperature sensors has a read-only property that outputs their sensor value. Both rooms host a heater ($H_A$ and $H_B$ in the figure) while the first room also hosts a cooler ($C$). Both heaters and the cooler have a single writable property that represents their power, which can be set manually to a value from 0 to 9 for the cooler, and on-off for the heaters.
Temperature values of both rooms are clearly affected by a number of factors: the power of the heaters, the power of the cooler, the temperature of the other room, and, finally, the temperature of the external environment. In the scenario described so far, we have $R = \{T_A, T_B, H_A, H_B, C\}$ and $W = \{H_A, H_B, C\}$, where $\{H_A, H_B, C\}$ are both readable and writable. The goal of the DTWT of such a complex system is to model the way in which these components affect each other, in order to output temperature values that resemble the real ones (i.e., $Z=\{T_A,T_B\}$).

Generally speaking, DTs are anticipated not only to model the devices' capabilities but also the communication occurring between each device or object within the network. In our case, this translates into the fact that the DT of a single entity should be able to consider and simulate the network communications involved among the IoT objects belonging to the single entity. Nevertheless, we highlight that the DT model created through the framework is strongly bound to the set of \textit{read-only} and \textit{writable} variables considered for each device. This means that the communications involved in the single entity can be modeled in the DT if and only if the communication capabilities of the devices are described through some \textit{read-only} and/or \textit{writable} properties, that is if the DT's intended use requires considering the communication as part of the Thing. In all the other cases, the communications among the IoT objects are considered granted and can be abstracted by the DT of the single entity, since they do not affect the other variables.

Clearly, it is not possible to predict the behavior of any IoT system without taking into account domain-specific information, since - for instance - the DT of a smart home may use ML techniques that require different information than the DT of an industrial appliance. 
For this reason, in this paper, we make use of ML with explicit models: each readable variable $r_j \in R$ is associated with a function that describes its behavior. We consider two types of functions in this study.
By relating some unknowns and their rate of change, \textit{differential equations} offer a powerful modeling tool to describe the evolution of physical quantities over time. Vice versa, \textit{algebraic expressions} allow to model the instantaneous value of specific quantities.
Once, combined, these two types of expressions allow to model a broad spectrum of cyber-physical systems and operations, capturing both physical and computational elements.

We highlight that the choice of using ML with explicit models rather than other popular approaches such as artificial neural networks is quite realistic since in several practical use cases (including the ones evaluated in Section~\ref{sec:validation}, and also industrial environments such as \cite{tila19}) the system behavior can be described with well-known models derived from the physics and the engineering.
As suggested in \cite{agrawal2022perspective}, a DT should impart some form of agency toward a \emph{desired function}, so the underlying model could be intended for a specific application and goal-oriented, capturing only some (cyber)physical aspects. Since all models are wrong, the question is whether the model is \emph{illuminating and useful} for the user \cite{box1976science}.
More general models could be considered as well, e.g., by including delayed or stochastic differential equations, or difference equations for discrete-time models. Physics-informed neural networks also give the opportunity to combine data-driven and model-based approaches.
All these components can be readily incorporated into our framework,
as such models build upon (known) equations and (partially unknown) parameters.
Then, one can capture features such as unobservable dynamics, response delays, and jitter in the communication network. A detailed discussion of these models is beyond the scope of this paper.

We split the set of readable variables $R$ into two disjoint sets $A = \{a_1, a_2, \dots\}$ and $B = \{b_1, b_2, \dots\}$, i.e., $R = A \cup B$, $A \cap B = \emptyset$, where $A$ contains the variables $r_j \in R$ that are modeled by an algebraic function and $B$ contains the variables modeled by a differential equation. 
%Let $A(t)$ and $B(t)$ be the value at time $t$ for all the state variables in $A$ and $B$, respectively.
In case the value of the variable $a_j \in A$ at time $t$ can be modeled by an algebraic expression $g_j$, then the latter is defined as follows:
\begin{equation}
\label{eq:algebraic}
a_j(t) = g_j(t, B(t), W(t), P_j) 
\end{equation}
%where $R_{-j} \equiv R \setminus \{r_j\}$ is the set of all the readable variable without $r_j$.
where $B(t)$ defines the value at time $t$ for all the state variables in $B$.
Here, $t$ is the time, $B(t)$ and $W(t)$ are the set of variables that $a_j$ could depend on, $P_j$ is the set of coefficients.
We define $G = \{g_j \,|\, a_j \in A\}$ as the set of the algebraic functions.

On the other side, in case the value of variable $r_j$ depends on its evolution over time, a differential equation is defined as follows:
\begin{equation}
\label{eq:differential}
\dot{b}_j(t) = f_j(t, B(t), A(t), W(t), P_j) 
\end{equation}
where $A(t)$ defines the value at time $t$ for all the state variables in $A$.
Similar to the previous case, $t$ is the time, $A(t)$, $B(t)$, and $W(t)$ are the set of variables that $b_j$ could depend on, and $P_j$ is the set of coefficients. %Obviously, being a differential equation, $f_j$ depends on $r_j(t)$.
Let $F = \{f_j \,|\, b_j \in B\}$ be the set of all the differential equations.

Both the algebraic and the differential equations defined for the variables $r_j$ depend on a set of model parameters $P_j$. We assume that $| P_j \bigcap P_{j'} | \geq 0$, $\forall r_j, r_{j'} \in R$, i.e., some model parameters can be shared among different state variables. We indicate with $P_{\mathrm{share}} = \bigcup_{r_j, r_{j'} \in R} \left( P_j \bigcap P_{j'} \right)$ the shared set of parameters.

The goal of the DT is to predict the system behavior by forecasting the values of the $Z$ variables at time $t'>t$, where $t$ is the actual time. To do that,
for each output variable in $Z$, the functions $F$ and $G$ are assumed to be known in the pre-deployment phase,  except for the exact values of the  model parameters $P$. The RDT framework allows to tune the value of the parameters
$P = \bigcup_{r_j \in R} P_j$
in order to properly estimate the values of the $Z$ variables, through the methods defined in Section~\ref{sec:learning}.
In the following, we indicate with $F^0$ and $G^0$ the \textit{behavioral models} where the parameters are initialized with the values defined by $P^0$, and with $\hat{F}$ and $\hat{G}$ the \textit{trained behavioral models}, i.e., the functions $\hat{F}$ and $\hat{G}$ in which all the parameters have been properly instantiated with the values $\hat{P}$ computed during the training phase.

In the context of the rooms scenario depicted in Figure~\ref{fig:roomscenario}, we partition the set of readable variables, denoted as $R$, into two subsets: $A=\{H_A, H_B\}$ and $B=\{T_A,T_B,C\}$. Variables in set $A$ require an algebraic expression for representation, whereas those in set $B$ necessitate a differential equation.
We emphasize the fact that the modeling decision is in charge of the user that is building the DT. Various configurations of sets $A$ and $B$ are feasible, and the subsequent definitions of algebraic and differential equations can vary accordingly.
For instance, the heater $H_A$ which belongs to set $A$, can be represented as an instantaneous value that characterizes the heater's power. This can be expressed as $H_A(t) = p_{H_A}$, where $p_{H_A} \in P_{H_A}$ represents the trainable parameter for $H_A$.
Conversely, the cooler $C$, a member of set $B$, can be described using a differential equation to capture the effect of its increasing power post activation. This can be given by $\dot{C}(t) = p_{C,1} \cdot ( p_{C,2} \cdot C_\text{ref}(t) - C(t) )$, where $C_\text{ref}$ is the cooler set-point power and $p_{C,1}, p_{C,2} \in P_{C}$ serve as the trainable parameters for $C$.
A comprehensive definition and further details of the rooms scenario will be elaborated upon in Section~\ref{ssec:validation-room}.

Given these premises, the key contributions of our work consist in how to:
\begin{itemize}
\item [(\textit{i})] allow the user to describe the characteristics of the IoT systems, such as the variables in $R$ and $W$, the IoT protocols to access them, and the model behavior $G$ and $F$, in a domain-independent way;
\item [(\textit{ii})] based on such inputs, fully automatize the data acquisition process from the IoT system in order to learn the values of the model parameters $P$ that best fit the observed data, and hence produce  the trained behavioral model ($\hat{F}$ and $\hat{G}$);
\item [(\textit{iii})] at the end of the learning phase, automatize the spawning of the DT which will consist in a software module enabling the user to interact with the virtual counterpart like with the original system.
\end{itemize}
Regarding the first contribution, our approach is to extend the WoT standard reviewed in the previous Section, and specifically the TD, with a new vocabulary that allows to model the DT of a generic IoT system and more specifically the functions $F$ and $G$ previously introduced. As shown in Figure~\ref{fig:scenario}, the IoT system ---considered as a single entity--- is mapped to one WT, exposing as properties the list of read-only and writable variables ($S$).
This is important, as canonical WoT deployments may also choose to map every IoT device to a WT, ending up with several WTs within a single environment. We instead propose to have a single WT where all IoT devices are mapped as affordances, as, otherwise, capturing their interactions would be far more challenging.
To this purpose, we map the read-only variables $R$ of the IoT devices to readable WoT Properties \texttt{(rProp)}, and the writable variables $W$ to writable WoT Properties \texttt{(wProp)}.
Notice that the choice of having a single WT to model the whole IoT environment is for readability; our proposal can support also scenarios with multiple interacting WTs, each associated with its own DT.

We remark that the  WoT is not a pre-requisite of our solution and that any IoT scenario that is not native WoT-based can be properly extended with WoT interfaces. However, the choice of the WoT is motivated by the possibility of describing ---in a uniform and domain-agnostic way--- the input/output and the behavior of the DT, thanks to our TD extension described in Section~\ref{sec:model}. The WT is then passed as input to the RDT framework, which is in charge of the next two contributions presented in Sections~\ref{sec:architecture} and \ref{sec:learning}, i.e., the generation of the trained behavioral model and then the spawning of the  Digital Twin-enabled Web Thing (DTWT). 
The latter has the same TD as the original WT, hence external services can interact with the DTWT in the same way as the original WT, by accessing the same list of affordances. However, any property change action generates a consequence only on the virtual environment without affecting the original system, hence enabling the what-if analysis. 
In addition, the user can modify a \texttt{time} property allowing to monitor the future behavior of the system, hence justifying the parallelism with the relativistic concept.
Additionally, the Thing Description of the W3C WoT provides a selection of well-established security mechanisms, along with metadata describing the configuration of these security mechanisms. These mechanisms and their configurations are directly built into the protocols eligible as Protocol Bindings for W3C WoT. This integration of security within the WoT framework ensures that our RDT framework can leverage these established protocols, thereby inheriting a robust security layer.

\section{W3C TD Extension} \label{sec:model}
In this Section, we discuss how to create the WoT TD of the environment, including its read-only and writable variables and the description of its behavioral model.
Notice that the system could have been represented by multiple WTs, one per IoT object, however, that would not allow us to use our framework seamlessly, because capturing dependencies among different WTs brings in additional challenges and needs more research in the future. In this paper, we use a single WT to represent a full IoT ecosystem, thus assuming that two different WTs do not explicitly influence each other, and a single WT can represent multiple IoT objects, with their affordances merged in a single pool.

Regarding the  structure of an ordinary \texttt{rProp} in the legacy W3C WoT standard, it includes the terms of the Thing Description (TD) Ontology\footnote{\url{https://www.w3.org/2019/wot/td}}: \texttt{type}, \texttt{description}, \texttt{observable}, \texttt{readOnly}, \texttt{writeOnly} and \texttt{uriVariables}.
We present below a brief recall:
\begin{itemize}
	\item \texttt{\textbf{type}} is a JSON-LD keyword to label the object with semantic tags about its type (e.g., ``number'').
	\item \texttt{\textbf{description}} is a human-readable piece of text that describes what the property is all about.
	\item \texttt{\textbf{observable}} is a boolean value that indicates whether the property is observable and thus if the servient hosting the WT needs to support the relative protocol binding.
	\item \texttt{\textbf{readOnly}} is a boolean value that indicates whether the property can only be read and not modified by a consumer.
	\item \texttt{\textbf{writeOnly}} is a boolean value that indicates whether the property can only be modified by a consumer and cannot be read.
	\item \texttt{\textbf{uriVariables}} is a data schema that indicates the list of parameters passed onto the request URI.
	\item \texttt{\textbf{forms}} is a list of hypermedia controls that represent how the interaction can be performed; for instance, in the case of an HTTP protocol binding, forms contain the endpoint and the port at which the property is accessible.
\end{itemize}
The behavioral model cannot be described in the current specifications of the W3C WoT standard. For this reason, in this paper, we propose to extend the existing TD with a new vocabulary allowing the general description of behavioral models for any domain, in order to support the next steps of automatic deployment of the DTWT.
Specifically, we added the terms \texttt{model}, \texttt{modelInput}, and \texttt{valueFrom} as shown in Table~\ref{table:terms} and  explained in the following. Since the vocabulary is not reviewed by the W3C WG yet, we use
the example namespace \texttt{http://example.org/2022/wot/dtwt}, so the recommended prefix to use in TDs is \texttt{dtwt:}.

\subsection{Model}
The field \texttt{model} is an arithmetical expression, encoded in a Python-like syntax, that describes the behavior of the property. This is basically represented via a function that is provided by the manufacturer of the appliance.
As described in Section~{\ref{sec:scenario}}, this field allows to describe the mathematical function that models the state of a readable variable $r_j \in R$.
Indeed, we  assume that the manufacturer knows such a function to some extent, obviously, the actual behavior of the appliance is determined by a number of factors that are dependent on the environment and are outside the control of the manufacturer a priori (i.e., the model parameters). In order to accurately represent this, the \texttt{model} is defined as a continuous function of a number of parameters that will be subsequently learned.
More formally, a model is defined in the following way:
\begin{lstlisting}[language=Json]
{behavior} = {function} | {constraints} | {guess}
\end{lstlisting}
Let us dissect the above syntax:

\subsubsection{behavior}
The behavior is one of \texttt{self} and \texttt{dot(self)}. In the first case, it means that the behavior of the property can be described by an algebraic expression, that is a fixed value that has no dynamics over time and depends on top of some static properties. An example might be the intrinsic power dissipated by a heater, which is dependent solely on its power level, manually set by the user.
Here, $r_j \in A$ will be described by a function in $G$ in the form of \eqref{eq:algebraic}.

In the second case, it means that the behavior of the property depends on time through a differential equation, therefore the function outputs its \textit{integration} over time. This includes properties that may change depending on the time passed since a certain event occurred, even if all the other properties of the DT remain constant.
An example can be the temperature of a room after the heater has been turned on. In fact, after such an event, the values do not change instantly, rather it takes a significant transient to reach a new steady state.
In this case, $r_j$ belongs to $B$ and the function is in the form of \eqref{eq:differential}.

\begin{table*}[t]
	\caption{DTWT terms defined for a Property. Their assignment is optional for TDs, but they are required if the Thing shall be turned into a DTWT.}
	\label{table:terms}
	\centering
	\begin{tabular}{ l p{30em} l l }
		\textbf{Term} (\texttt{dtwt:}) & \textbf{Description} & \textbf{Class} & \textbf{Type} \\ 
		\hline
		\texttt{model} & Mathematical model of the property & PropertyAffordance & string \\
		\texttt{modelInput} & Container for inputs of the model & PropertyAffordance & Class \\
		\texttt{ValueFrom} & Indication from where to take the value of the property, i.e., from the physical sensor or from the computed model & PropertyAffordance & string
	\end{tabular}
\end{table*}

\subsubsection{function}
The function is an arithmetical expression, encoded in python-like syntax. Apart from common numbers, operands, and parentheses, we have added a few keywords that point to definite values in our TD. 
First of all, the function is defined on top of the set $P$ of parameters that, as said, are tuned by the optimizer to fit the model to the real behavior of the property. These parameters are organized in two arrays:
\begin{itemize}
	\item \texttt{params[n]} is an array of $n$ parameters that are specific to this model (i.e., their scope is limited to this function). Any parameter can be accessed using \texttt{params[i]} where $0\leq i < n$.
	\item \texttt{global[m]} is an array of $m$ parameters that are global and defined only once (i.e., their scope is extended to the whole thing description). Any parameter can be accessed using \texttt{global[i]} where $0\leq i < m$.
\end{itemize}
The distinction between \texttt{global} and \texttt{local} parameter is needed when this latter is shared among multiple state variables. For this reason, the \texttt{global[m]} array will contain the whole set $P_{\mathrm{share}}$.
Furthermore, the function can use other properties of the same DTWT, which are called ``model inputs'' and are listed in the \texttt{modelInput} field of the \texttt{rProp}.
An example may be our temperature property that takes as input the power of the heater, which is another \texttt{rProp} of the DTWT. The detailed interaction between a model input and the model that imports it is described in the next subsection.
Finally, the function can use the keyword \texttt{self}, which represents the non-integrated value of the property, and the keyword \texttt{value()}, which represents the actual value read from the real sensor, when available. 

\subsubsection{constraints}
The \textit{constraints} is a set of preconditions regarding the \texttt{params[n]} and the \texttt{global[m]} of the model. They are read by the optimizer and respected when performing the parameter optimization.
We assume that each parameter $p_j \in P$ is bounded by a minimum and a maximum value, i.e., $p_j^{\mathrm{min}} \leq p_j \leq p_j^{\mathrm{max}}$. The set of all the parameter constraints forms the hyperbox $\setP$.
For instance a valid syntax can be \texttt{params[0] >= 0.0}, \texttt{params[1] <= 0.0}, \texttt{global[0] <= 1.0}.
The global parameters' constraints are to be set only once throughout the TD (clashing constraints will lead to errors).

\subsubsection{guess} The \textit{guess} is a set of initial value assignment regarding the \texttt{params[n]} and the \texttt{global[m]} of the model, defining the set $P^0$. They are the default values before the optimization takes place. For instance, a valid syntax can be \texttt{params[0] = 2.5}, \texttt{params[1] = 0.0}, \texttt{global[0] = 0.01}.
The global parameters' guess, as for the constraints, is to be set only once throughout the TD (clashing assignments will lead to errors).

\subsection{Model Inputs}
The field \texttt{modelInputs} is a list of other \texttt{rProps} of the WT that are used within the model. Each of these has five fields: 
\begin{itemize}
	\item \texttt{title}, which is a friendly name, 
	\item \texttt{propertyName}, which is the exact name of the \texttt{rProp} of the WT that this input is referring to, 
	\item \texttt{type}, which is the JSON-LD type, 
	\item \texttt{model}, which is again a model (using the same syntax as described above) that defines how this input affects the model of the property, and 
	\item \texttt{modelType}, which is an optional field denoting a class assigned to this input that is only valid within the scope of the property and is used to aggregate all inputs belonging to the same \texttt{modelType} (if there is more than one).
\end{itemize}

The property \texttt{model} can use the \texttt{modelInputs} in two ways. 
The first is by using the construct ``\texttt{input(\{title\})}''; this imports the exact model of the \texttt{modelInput} identified by ``\texttt{title}''. The model input may have its own \texttt{params[]}: in this case, \texttt{params[i]} of the property model is not equal to \texttt{params[i]} of the model input, as they have different scoping. The model input may refer  to the value outputted by the actual property identified by \texttt{propertyName} though the keyword \texttt{self}.
Alternatively, the property \texttt{model} can import a group of model inputs through the construct \texttt{inputType(\{modelType\})}; in this case, the property model imports all the model inputs with the type \texttt{modelType} and aggregates them. The aggregation takes place through an aggregation function which must enclose the \texttt{inputType()} call (for instance \texttt{sum(inputType({modelType}))}).

\subsection {\textit{ValueFrom}}
The field \texttt{valueFrom} describes whether there is a real property that can be read (e.g., a real sensor), in such case the value of this field will be set to ``readProperty''. Alternatively, the field can be set to ``model'', which instead triggers the simulation of the property value (depending on the behavior, via algebraic evaluation or integration over time) as described by the model field of the \texttt{rProp}. This indicates where the data comes from when querying directly the \texttt{rProp} via its URI, however, when the latter gets imported into a model, it is its model that always gets imported.

In the Appendix, we show the JSON code of the TD of the smart home environment, restricting our attention to the new fields which are used to describe the behavioral model of the IoT system.

\section{Framework Architecture and Implementation}\label{sec:architecture}
\begin{figure}
	\centering
	\includegraphics[width=0.49\textwidth]{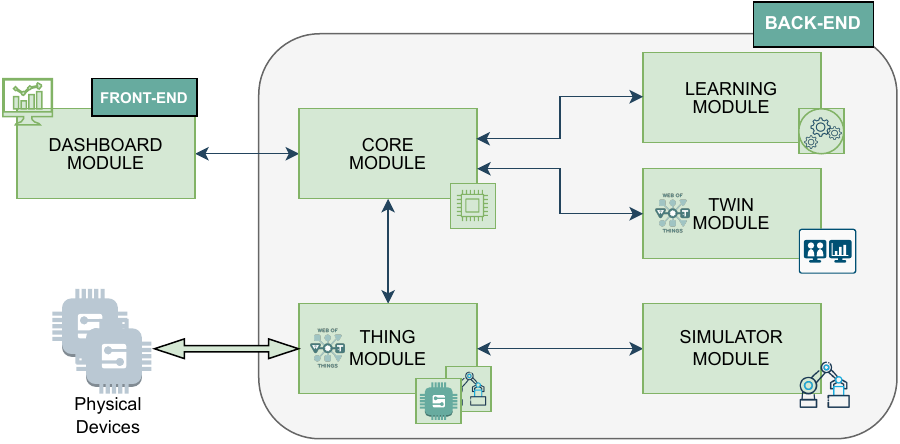}
	\caption{The RDT framework architecture.}
	\label{fig:architecture}
\end{figure}

\begin{figure*}
	\centering
	\includegraphics[width=0.9\textwidth]{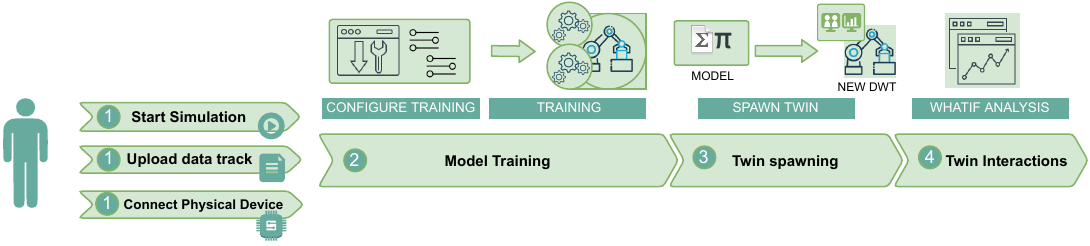}
	\caption{The sequence of operations of the RDT framework. The user can interact with it in three ways: connecting it to a simulator, uploading tracks with property values, or connecting it to a physical device. }
	\label{fig:interaction_flow}
\end{figure*}

\begin{figure}
	\centering
	\includegraphics[width=0.98\linewidth]{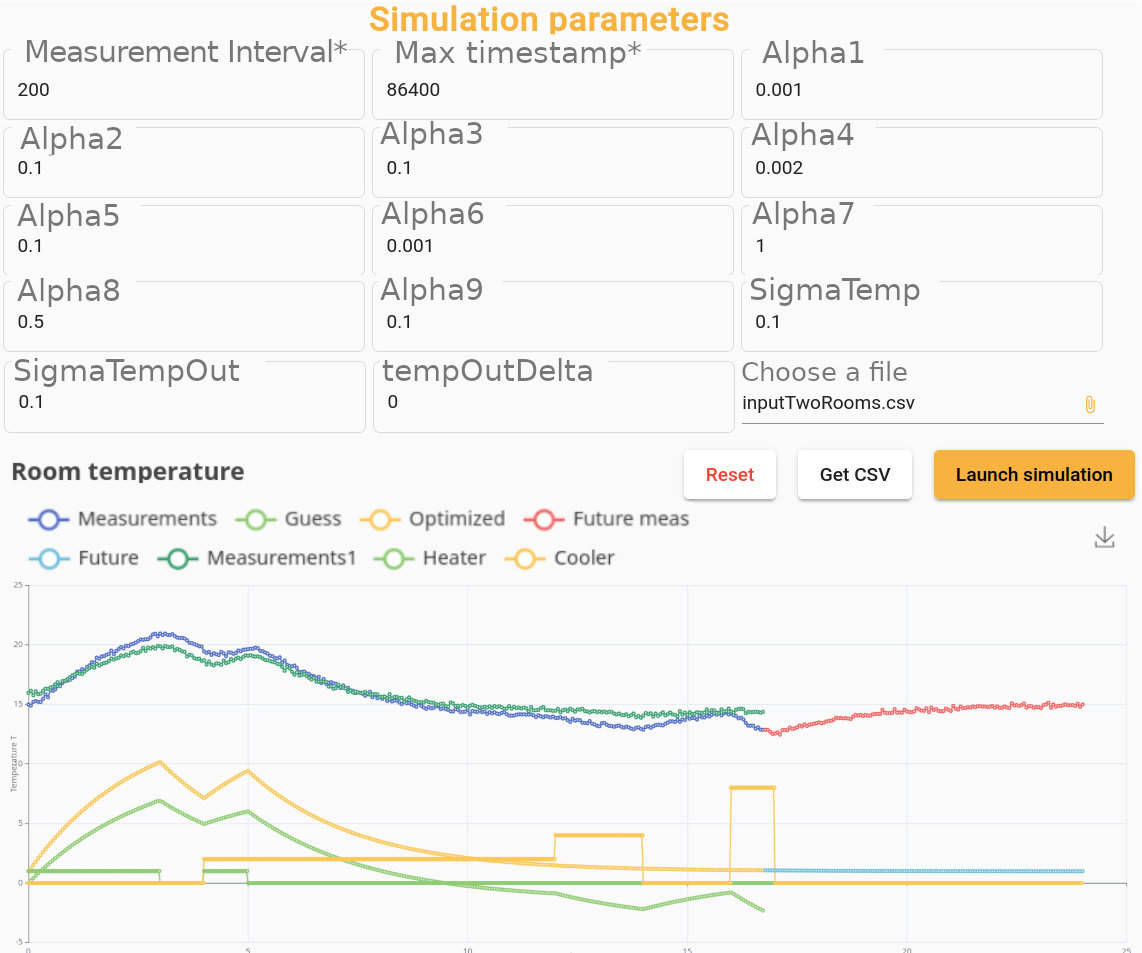}
	\caption{A screenshot of the \textit{dashboard module}. Using such GUI, the user can simulate, train, and visualize the guess of the model of the room scenario described in Section~\ref{sec:model}.}
	\label{fig:dashboard}
\end{figure}

Based on the TD extension described in the previous Section, we designed the RDT framework to support the automatic generation of a DTWT. The framework allows to monitor the properties of the WT, to store their values, to fit the behavioral model till creating the trained model, and, finally, to generate the DTWT. The RDT framework architecture is micro-services oriented, with each macro-functionality mapped to a specific module. As shown in Figure~\ref{fig:architecture}, the architecture is composed of six main modules: a single one for the \textit{front-end} functionalities, while the other five run on the \textit{back-end}. More in detail, the \textit{Dashboard module} provides an intuitive Graphical User Interface (GUI) through which users can easily interact with the framework.

We detail a typical flow of operations in Figure~\ref{fig:interaction_flow}.  Three main interaction modes are supported:
simulating the IoT system operations and producing simulated sensory data, uploading traces containing the measurements, or  attaching a physical device to the platform via the API of the RDT framework.
Using any of the previous options, the user provides the training data for the parameters' identification and then configures the training hyperparameters, such as the tolerance for termination of the training algorithm, or specifies the algorithm itself.
Once the model has been trained, a new DTWT sharing the same TD as the original WT can be spawned; the DTWT is completely detached from the physical IoT device and hence can be used for what-if analysis (i.e., the user can observe how the twin would react to future events). In addition, the framework supports re-sync operations when the properties of the DTWT can be re-initialized with the current values of the original WT.
The back-end functionalities are implemented within (i) the \textit{core module}, (ii) the \textit{learning module}, (iii) the \textit{twin module}, (iv) the \textit{Thing module}, and (v) the \textit{simulator module}. The \textit{core module} enables the communication between the \textit{dashboard} and the rest of the services through a dedicated API, by translating the user's requests into  proper commands to be invoked on the services. Additionally, the \textit{core module} monitors the \textit{Things}, collecting all the measurements and data needed for the \textit{training phase} from the \textit{learning module}. The latter, as better detailed in Section~\ref{sec:learning}, allows to parse the TD and to instantiate the behavioral model described by our TD extension. Furthermore, it uses the data provided by the \textit{core module}  to derive the optimal value of the model parameters.
Within the RDT framework, the data storage is primarily dependent on the specific application and it is not directly related to the framework itself. It depends on the amount of data collected for training the behavioral models. Different applications may generate varying volumes of data, which in turn influences the storage requirements. Our framework is designed to be adaptable to these varying needs, ensuring that the data storage solutions can be tailored according to the specific demands of each application.
The \textit{twin module} takes in input the trained model and generates the DTWT, 
which forecasts the values of the properties 
through the trained behavioral model.
The \textit{Thing module} contains all the mechanisms to interact with the IoT counterpart. This can be a physical device --- in this case, a dedicated channel is created for the data exchange --- or can be simulated through the \textit{simulator module}.
The latter is just a wrapper for domain-specific tools that allow reproducing the behavior of a device, by generating the values of its properties over time and allowing to receive actions in input. As a result, any custom simulator can be connected to the RDT framework by implementing a proper connector.
Clearly, the process of training, learning, spawning, and simulating may be repeated over time, as the behavior of the physical devices may change or there may be aspects that had never manifested during the previous training iteration. In such case, the newly generated DTWTs will replace the old ones.
The full process designed within the RDT framework primarily introduces overhead during the automatic generation and deployment of digital twins. However, it is crucial to emphasize that instantiating new digital twins within our framework requires constant resources. This efficiency in spawning new digital twins ensures that the framework can scale effectively, accommodating the creation and management of multiple digital twins without significant increases in computational load or complexity. Such scalability is a key feature of our framework, enabling it to adapt and respond to varying demands in diverse IoT environments.
The most time-consuming aspect of the RDT framework is the training phase. The real-time efficiency in the re-spawning phase is influenced by the specific problem and the complexity of the training model. While our framework is designed for efficient re-training process, the actual speed and responsiveness in real-time scenarios depend on the training model's intricacies and the data nature. Therefore, the RDT framework's effectiveness in real-time operations varies with each application's specific requirements and complexity.

\subsection{Implementation}
The front-end functionalities are included in the \textit{Dashboard module} which consists of a Web application written using the \texttt{Angular}\footnote{\url{https://angular.io/}} framework, as shown in Figure~\ref{fig:dashboard}; the \texttt{Apache} \texttt{Echarts}\footnote{\url{https://echarts.apache.org/en/index.html}} library is used to plot the results of the what-if analysis, i.e., showing the real measurements and the predictions made by the DT. The back-end functionalities are implemented within the other five modules of the architecture: each of them is implemented in \texttt{Typescript} and runs over \texttt{NodeJS} as runtime system. The \textit{Thing} and the \textit{twin} modules have been implemented as WTs using the \texttt{Eclipse} \texttt{node-wot} tool\footnote{\url{https://github.com/eclipse/thingweb.node-wot}}. More in detail, each WT runs a \texttt{Servient}, as indicated in the official guidelines of the W3C WoT Scripting API\footnote{\url{https://www.w3.org/TR/wot-scripting-api/}}. The \textit{learning module}  is a Web server that collects requests from the core module through a REST API. The optimization tasks are  implemented as Python scripts using the \texttt{SciPy} library, as better detailed in Section~\ref{sec:learning}. The \textit{simulator module} is written in \texttt{Typescript}.

\section{DT Learning}\label{sec:learning}
The \textit{learning module} is a crucial component of the RDT framework. It is in charge of parsing the TD of the WT, extracting its behavioral model, and generating the trained behavioral model from the data.
The operations of the module can be split into three major tasks: (\textit{i}) model import, (\textit{ii}) model inference, and (\textit{iii}) model learning. In the following, a detailed description of each  task is provided.

\subsection{Model Import}\label{ssec:modelimport}

A \textit{parser} module has been implemented in order to translate the arithmetical expressions that are encoded in a Python-like syntax within the TD of the WT, i.e., the algebraic functions $G$ and the differential equations $F$, into an executable Python code.
Listing~\ref{lst:td-temp-parsedmodel} shows an example of the translation of the temperature property for the smart home example (Listing~\ref{lst:td-temp}).

\begin{lstlisting}[caption={DTWT TD, parsed room model}, label=lst:td-temp-parsedmodel, language=Python]
y[0]=readProperty("heater",timestamp,data)
dxdt[0]=params[1]*(params[2]*(params[3]-x[0])+((params[0]*y[0])+(-x[2]))+params[4]*(x[1]-x[0]))
dxdt[1]=params[6]*(params[7]*(params[3]-x[1])+((params[5]*y[0]))+params[4]*(x[0]-x[1]))
dxdt[2]=params[8]*(params[9]*max(0,min(round(readProperty("cooler",timestamp,data)),9))-x[2])
\end{lstlisting}

In Listing~\ref{lst:td-temp-parsedmodel} we can notice the differences between the algebraic variables (the \texttt{y}) and the differential properties (the \texttt{x} and the \texttt{dxdt}). 
In this specific example, the variables \texttt{x[0]} and \texttt{x[1]} correspond to the temperature values of $R_A$ and $R_B$ at time \texttt{timestamp}, while \texttt{y[0]} and \texttt{x[2]} correspond to the heating and cooling power, respectively. The rationale of the equations is discussed in Section~\ref{ssec:validation-room}.

\subsection{Model Inference}\label{ssec:modelprediction}
The \textit{inference} task returns the history of states in a specific time interval, given the initial state, the model parameters and the control actions; i.e., it is used as a mathematical simulation engine to build past or future states.
More in detail, given $t$ in a time interval $\setT_T \coloneqq [t_0, t_T]$ from an initial time $t_0$ to an end time $t_T$, 
the actions executed on the writable properties performed during this time interval $W_{T}$, some model parameters $\widehat{P}$, and an initial state $\widehat{R}(t_0)$, the models defined by the sets $F$ and $G$ (in the form of the Equations \eqref{eq:algebraic}-\eqref{eq:differential}) allow the prediction of the states $R(t)$ over $\setT_T$.
To this aim, we introduce a parametric and non-autonomous ordinary differential equation (ODE) that is defined for each $b_j \in B$ from the executed actions and the model parameters:
\begin{align}
	\dot{b}_j(t) &= \mathcal{F}_{j,W_{T},\widehat{P}}(t, B(t)) \label{eq:cauchy-new} \\
	&= f_j(t, B(t), A(t), W_{T}(t), \widehat{P}_j), \quad t\in \setT_T , \nonumber\\
	r_j(t_0) &= \widehat{r}_j(t_0). \nonumber
\end{align}
Solving the Cauchy -- or initial value -- problem results in a prediction of the DTWT behavior according to the parametrized model.
In this work, we employed ODE solvers based on Runge-Kutta schemes \cite{dormand1980family}, \cite[Ch. 5]{gerdts2011optimal}, which provide a numerical approximation $\widehat{b}_j(t)$ of a solution to \eqref{eq:cauchy-new}.
Then, one can readily compute the approximation of $\widehat{a}_k \in A$ of the algebraic states using \eqref{eq:algebraic} pointwise.
For our purposes\footnote{In our specific implementation, we use the function \texttt{solve\_ivp} from the SciPy library \texttt{scipy.integrate} \cite{scipy2020}. A modern ecosystem for solving differential equations is described in \cite{rackauckas2017differentialequations}.}, the ODE solver provides
$B_{W_{T}, \widehat{P}, \widehat{R}(t_0)}(t)$ and $A_{W_{T}, \widehat{P}, \widehat{R}(t_0)}(t)$, namely it returns
the trajectories of any $b_j \in B$ and $a_k \in A$ on a time interval, given a history of actions $W_{T}$, a set of parameters $\widehat{P}$ and an initial state $\widehat{R}(t_0)$.

\subsection{Model Learning}\label{ssec:modellearning}
The inference task assumes the knowledge of the model parameters $P$ (called $\widehat{P}$ in Section~\ref{ssec:modelprediction}).
However, these values are not known a priori and need to be estimated. For this reason, 
this task is in charge of estimating $P$ with $\widehat{P}$ starting from the observed behavior of the WT.
Let $\{ \Omega_1,\ldots,\Omega_O \}$ be the set of $O$ observations of the WT at the time instants $\{t_1,\ldots,t_O\} \subset \R$.
Let the function $h(\cdot)$ be an observation model, such that
\begin{equation*}
	\Omega_k \approx h(t_k, B(t_k), A(t_k), W(t_k), \widehat{P})
\end{equation*}
for $k=1,\ldots,O$.
Effectively, given some past actions, we seek the optimal choice of parameters $\widehat{P}$ that minimizes the deviation between observations and retrospective predictions based on the parametric model.

Given a set of actions $W_{Q}$ executed in $\setT_Q \coloneqq [t_0, t_Q]$, with $t_Q \geq t_O$, and
defining the predictor
\begin{equation*}
	H_{W_{Q}, \widehat{P}, R(t_0)}(t)
	\coloneqq
	h(t, B_{W_{Q}, \widehat{P}, R(t_0)}(t), A_{W_{Q}, \widehat{P}, R(t_0)}(t), W_{Q}(t), \widehat{P}) ,
\end{equation*}
we consider the following optimization problem to fit the timestamped observations $(t_k,\Omega_k)$, $k=1,\ldots,O$:
\begin{equation}
	\label{eq:nlsq}
	\minimize_{\widehat{P}\in\setP ,\, R(t_0)\in\setR_0}
	\quad
	\sum_{k=1}^O \| H_{W_{Q}, \widehat{P}, R(t_0)}(t_k) - \Omega_k \|^2 .
\end{equation}
Here, $\setP$ is the hyperbox containing the lower and upper bounds of each parameter,
and $\setR_0$ is the set of the possible initial states.
In the smart room example presented so far, $\setR_0$ is composed of the constraints: $-20 \leq T_A(t_0), T_B(t_0) \leq 50$, $0 \leq H_A(t_0), H_B(t_0) \leq 1$, and $0 \leq C(t_0) \leq 9$.

The resulting mathematical task \eqref{eq:nlsq} is a nonlinear least squares problem with bound constraints \cite[Ch. 4]{nocedal2006numerical}.\footnote{In our specific implementation, we use the function \texttt{least\_squares} from the SciPy library \texttt{scipy.optimize} \cite{scipy2020}.}
Although in general nonconvex and possibly with many local minimizers, the complexity of algorithms for tackling problem \eqref{eq:nlsq} is addressed in \cite{cartis2022evaluation}, discussing how many function evaluations might be required to obtain an approximate solution.

Starting from some guesses $P^\guess$ and $R^\guess(t_0)$, and seeking some optimal values $P^\star$ and $R^\star(t_0)$, the numerical solver needs to compute $H_{W_{Q},\widehat{P},\widehat{R}(t_0)}(t_k)$ for some tentative values $\widehat{P}\in\setP$ and $\widehat{R}_0(t_0) \in \setR_0$.
In turn, such evaluation requires to solve an initial value problem as in \eqref{eq:cauchy-new}, obtaining $B_{W_{Q},\widehat{P},\widehat{R}(t_0)}(\cdot)$ and $A_{W_{Q},\widehat{P},\widehat{R}(t_0)}(\cdot)$.
Then, these predicted trajectories are sampled at the time instants $t_k$, $k=1,\ldots,O$, where the measurement model $h$ is evaluated, yielding $H_{W_{Q},\widehat{P},\widehat{R}(t_0)}(t_k)$.

Given the estimates, $\widehat{P}$ and $\widehat{R}(t_0)$ of model parameters and (differential) state at time $t_0$, the \textit{learning module} is able to predict the DTWT state at any time for some given history of actions on the writable variables $W_{Q}$, effectively simulating the system dynamics and behavior according to the identified model. In this case, the $W_{Q}$ function is needed to simulate the action executed in future times $t_k$, i.e., for $t_O < t_k \leq t_Q$.

\section{Validation} \label{sec:validation}
In this Section, we present a comprehensive analysis and validation of the proposed RDT framework.
We assess the performance on two different case studies:
(\textit{i}) a simulation test where we present the results of the smart home scenario introduced firstly in Section~\ref{sec:model},
and (\textit{ii}) a real test-bed consisting of a drone quadcopter which can be employed e.g., for video-surveillance applications.
In the following two subsections, we  describe the setup of the experiments and show the evaluation results.

\subsection{Simulated testbed: smart home} \label{ssec:validation-room}

In Section~\ref{sec:model} we introduced the smart home environment  as a showcase of the WT TD extension.
In this Section, we detail the mathematical model characterizing the system behavior and we evaluate the performance of the learning process.

The proposed scenario is composed of two closed rooms: room $R_A$ and room $R_B$.
At time $t$ the temperature of $R_A$ is denoted by $T_A(t)$, the temperature of $R_B$ by $T_B(t)$, and the outdoor temperature by $T_\text{out}(t)$. 
Inside each room (differently from our initial example) we have a heater that we assume to be connected to the same heating system, i.e., they are simultaneously either \textit{ON} or \textit{OFF}.
The heaters' status at time $t$ is defined by the binary variable $b_H(t)$: 
\begin{equation*}
	b_H(t) = \begin{cases}
		1 & \text{if the heating system is \textit{ON}} \\
		0 & \text{otherwise}
	\end{cases}
\end{equation*}
The heaters' power at time $t$ are defined by $H_A(t)$ and $H_B(t)$ for the heater in $R_A$ and $R_B$, respectively.
In $R_A$ we suppose to have also a cooler whose power at time $t$ is denoted by $C(t)$; this depends on a user-defined setpoint $C_{\mathrm{ref}}(t)$ which has $10$ degrees of active powers, namely $0 \leq C_{\mathrm{ref}}(t) \leq 9$, $C_{\mathrm{ref}}(t) \in \mathbb{N}$ for all times $t$.
Based on our previous notation, we have the set of readable variables $R =$ $\{T_A,$ $T_B,$ $H_A,$ $H_B,$ $C,$ $C_{\mathrm{ref}},$ $b_H\}$ and writable variables $W =$ $\{C_{\mathrm{ref}},$ $b_H\}$.

The following differential equations describe the system dynamics over time $t$; they will be used in the TD of the WT:
\begin{subequations}\label{eq:roomModel}
	\begin{align}
		\dot{T}_A(t) &= \alpha^{R}_1 [ \alpha^{R}_2 (\alpha^{R}_6 - T_A(t)) + H_A(t) - C(t) + \nonumber \\
		& \quad + \alpha^{R}_3 (T_B(t) - T_A(t)) ] \label{eq:roomModelFirst}\\
		\dot{T}_B(t) &= \alpha^{R}_4 [ \alpha^{R}_5 (\alpha^{R}_6 - T_B(t)) + H_B(t) + \nonumber \\ 
		& \quad + \alpha^{R}_3 (T_A(t) - T_B(t)) ] \\
		H_A(t) &= \alpha^{R}_7 b_H(t) \\
		H_B(t) &= \alpha^{R}_8 b_H(t) \\
		\dot{C}(t) &= \alpha^{R}_9 \left[ \alpha^{R}_{10} C_\text{ref}(t) - C(t) \right]  \label{eq:roomModelLast}
	\end{align}
\end{subequations}
Herein, the linear model for the temperature dynamics is provided by Newton's law of heat transfer, whereas considering thermal inertia in the cooling system yields the first-order linear model in \eqref{eq:roomModelLast}. 
The parameters set $P$ is defined by $P = \{\alpha^{R}_1, \dots, \alpha^{R}_{10}\}$, where 
$\alpha^{R}_*$ are free variables and their values will be estimated by the learning module of the RDT framework, based on observations from the simulated/real environment.

Given such a model, we generated the TD for the rooms as previously introduced in Section~\ref{sec:model}.
In this description, we note that some parameters are shared among different properties.
For example, $\alpha^R_3$ is connected to both $R_A$ and $R_B$.
The parameter $\alpha^R_6$ is shared between the two heaters but it coincides with the outdoor temperature, assumed to be constant over time, i.e., $T_{\text{out}}(t) = T_{\text{out}} = \alpha^R_6$, $\forall t$.
Finally, we defined a \texttt{global} variable for the outdoor temperature, called \texttt{global[2]} in Listing~\ref{lst:td-temp}.

We set up a simulation environment to generate the temperature data acquired by the virtual sensors positioned inside the two rooms.
To avoid a straightforward learning system where the proposed DTWT model mimics exactly the simulated environment, we defined a different model to simulate the rooms' temperature.
Considering a thermal inertia, and thus first-order dynamics, in the heating system and not in the cooling system, the simulation model reads as follows:
\begin{subequations}\label{eq:roomSimulation}
	\begin{align}
		\dot{T}_A(t) &= \beta_1 [ \beta_2 (T_\text{out}(t) - T_A(t)) + H_A(t) - C(t) + \nonumber \\
		& \quad + \beta_3 (T_B(t) - T_A(t)) ]  \label{eq:roomSimFirst} \\
		\dot{T}_B(t) &= \beta_4 [ \beta_5 (T_\text{out}(t) - T_B(t)) + H_B(t) + \nonumber \\ 
		& \quad + \beta_3 (T_A(t) - T_B(t)) ] \\
		\dot{H}_A(t) &= \beta_6 [\beta_7 b_H(t) - H_A(t) ] \\
		\dot{H}_B(t) &= \beta_6 [\beta_8 b_H(t) - H_B(t) ] \\
		C(t) &= \beta_9 C_{\text{ref}}(t)  \label{eq:roomSimLast}
	\end{align}
\end{subequations}
Parameters $\beta_*$ in \eqref{eq:roomSimulation} are used to customize the rooms' simulations and are unknown to the user. Where not specified otherwise, their default values are indicated in Table~\ref{tab:evalpar}.

\begin{table}[!b]
	\centering
	\caption{Parameters used in the simulated room environment}
	\label{tab:evalpar}
	\begin{tabular}{cc|cc|cc}    
		\hline                     
		$\beta_1$ & $\SI{0.001}{K/J}$ & $\beta_4$ & $\SI{0.002}{K/J}$ & $\beta_7$ & $\SI{1.0}{W}$ \\
		$\beta_2$ & $\SI{0.1}{W/K}$ & $\beta_5$ & $\SI{0.1}{W/K}$ & $\beta_8$ & $\SI{0.5}{W}$ \\
		$\beta_3$ & $\SI{0.1}{W/K}$ & $\beta_6$ & $\SI{0.001}{s^{-1}}$ & $\beta_9$ & $\SI{0.1}{W}$ \\
		\hline                
	\end{tabular}
\end{table}

The observed parameter values at time $t_k$ are represented by the set $\Omega_k =\{T_A^{\text{obs}}(t_k), T_B^{\text{obs}}(t_k), H_A^{\text{obs}}(t_k), H_B^{\text{obs}}(t_k), C^{\text{obs}}(t_k), b_H^{\text{obs}}(t_k)\}$.
To make the simulation more realistic, we added a Gaussian noise to the observed temperature  during the learning process, i.e.,
\begin{align}
	T_A^{\text{obs}}(t_k) &= T_A(t_k) + \varepsilon \\
	T_B^{\text{obs}}(t_k) &= T_B(t_k) + \varepsilon \label{eq:obsSecond}
\end{align}
where $\varepsilon \sim \mathcal{N}(\mu,\sigma)$ simulates the error introduced by the devices.
Finally, also in the simulation part, we keep the $T_{\text{out}}$ constant with an additive Gaussian noise, i.e., $T_{\text{out}} = \SI{15}{\degreeCelsius} + \mathcal{N}(\mu,\sigma)$.
In the following tests, we used $\mu = \SI{0}{K}$ and $\sigma = \SI{0.1}{K}$.

\begin{figure*}%
	\centering
	\subfloat[]{\includegraphics[width=0.33\textwidth]{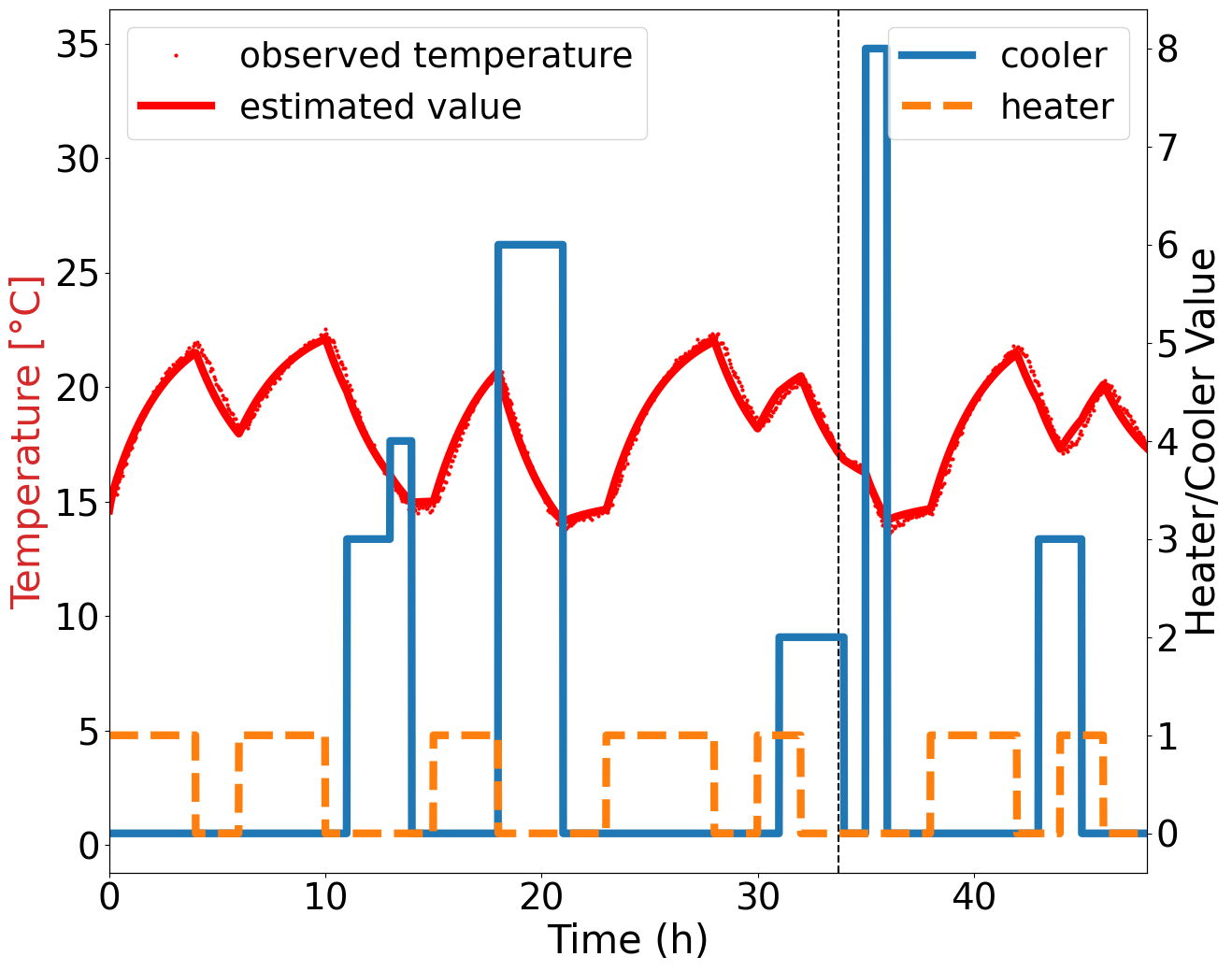} \label{fig:roomeval1}}
	\subfloat[]{\includegraphics[width=0.33\textwidth]{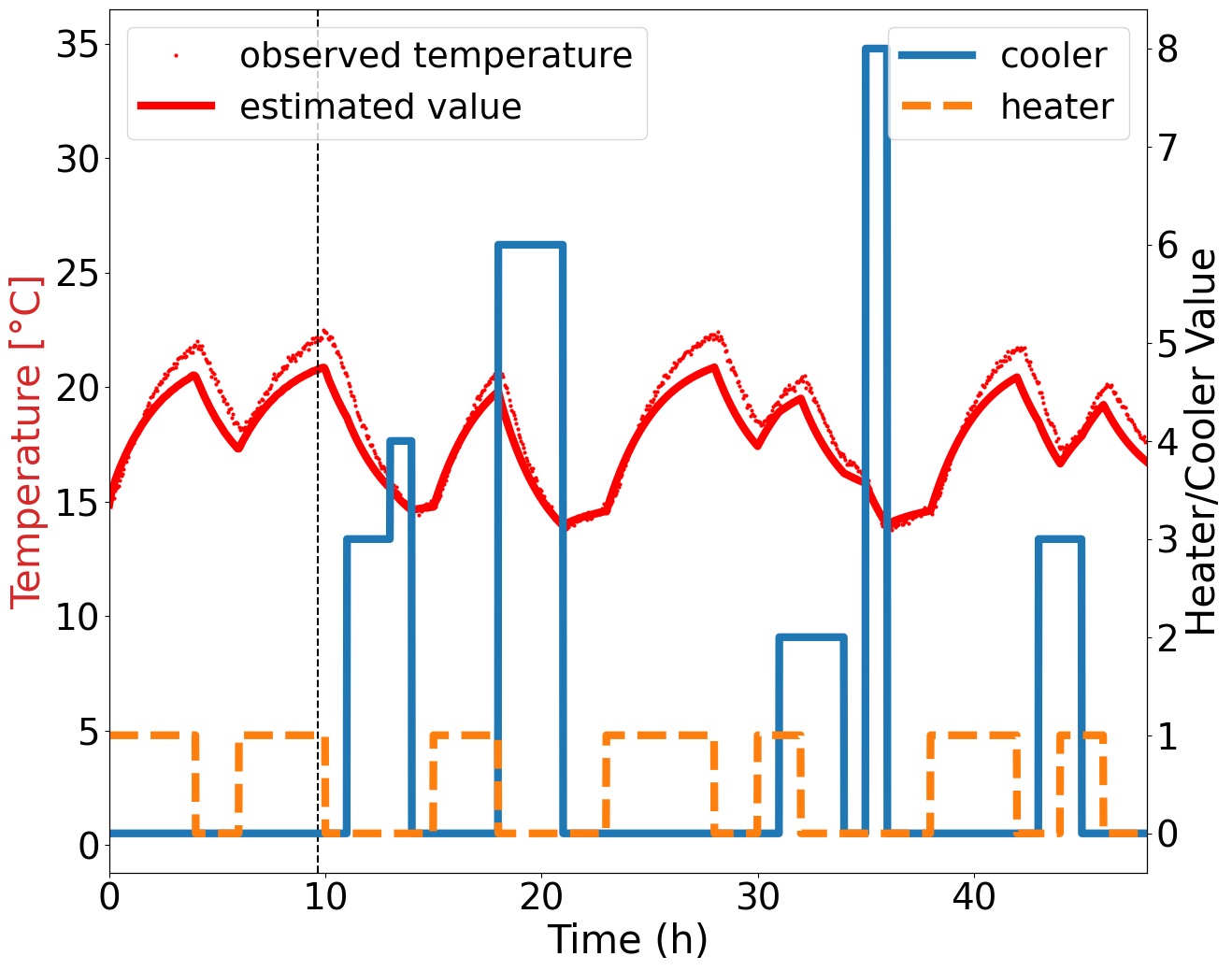} \label{fig:roomeval2}}
	\subfloat[]{\includegraphics[width=0.33\textwidth]{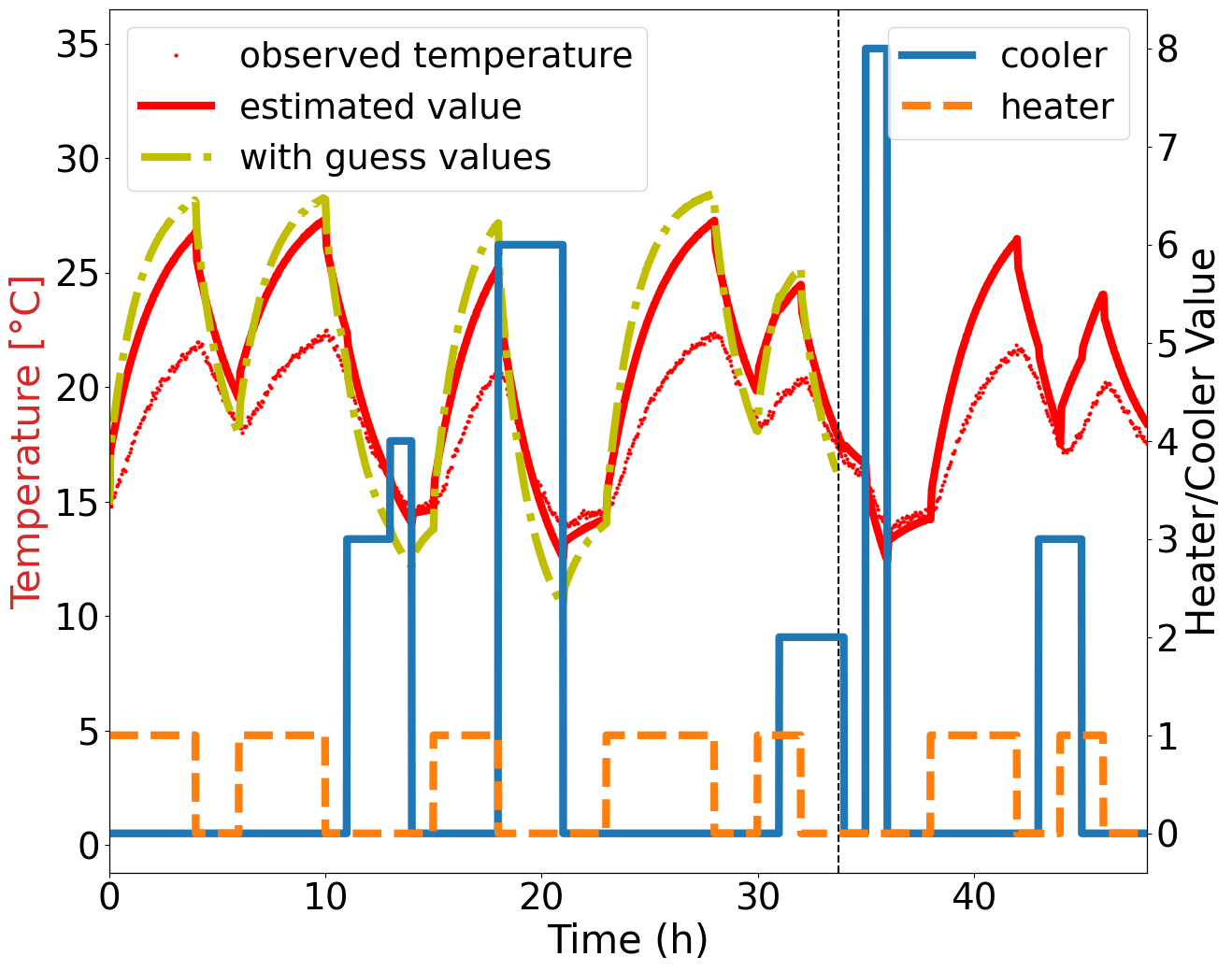} \label{fig:roomeval3}}
	\caption{The evaluation of the accuracy of $T_A(t)$ estimated by the spawned DTWTs. Figure~\ref{fig:roomeval1} shows the observed and estimated values when the DTWT is spawned at time \SI{34}{h}; Figure~\ref{fig:roomeval2} shows the same curves when the DTWT is spawned at time \SI{10}{h}; Figure~\ref{fig:roomeval3} shows the same results when the DTWT is spawned at time \SI{34}{h} but without any initial calibration of the parameter guesses $p^\guess$.}
	\label{fig:roomeval}
\end{figure*}

In the experiments, we evaluated the ability of the trained behavioral model to forecast indoor temperature.
We generated two DTWTs for the two rooms, and we set up the framework in order to learn the value of $T_A$.
As described in Figure~\ref{fig:interaction_flow}, we  uploaded the tracks containing the temperature measurements from time $t=0$ to the time of the DTWT generation $t_{\text{spawn}}$.
These tracks are generated using the rooms simulator described by Equations~\eqref{eq:roomSimFirst}-\eqref{eq:obsSecond} generating the observations $\Omega_k$, for $0 < k < t_{\text{spawn}}$, using a predefined sequence of actions $W_{\Gamma_{\text{spawn}}}$ for the writable variables (shown in blue and orange lines in Figures~\ref{fig:roomeval1}-\ref{fig:roomeval3}).
Figures~\ref{fig:roomeval1}-\ref{fig:roomeval3} show the observed and forecast values of $T_A$ as produced by the DTWT, for different $t_{\text{spawn}}$ values, i.e., for different training lengths.
The observed values are depicted with a dotted red line, and the forecast values with a solid red line; similarly, we highlight the event of training end at time $t_{\text{spawn}}$ with a vertical dashed black line.
We set $t_{\text{spawn}} = \SI{34}{h}$ for Figures~\ref{fig:roomeval1} and \ref{fig:roomeval3}, and $t_{\text{spawn}} = \SI{10}{h}$ for Figure~\ref{fig:roomeval2}.
In all the Figures, we show the operational state of the heater and cooler over time, i.e., their power values, tuned as a consequence of the user's actions.
For the experiment in Figure~\ref{fig:roomeval3}, the parameters $p^\guess$ were initialized with random values.

Figure~\ref{fig:roomeval1} shows that the proposed model is able to predict the room temperature in a very accurate way.  Indeed, after the training interval (equal to \SI{34}{h}), the line of the estimated values overlaps with the line of the observed values.
In Figure~\ref{fig:roomeval2} we reduced the training length compared to the previous analysis ($t_{\text{spawn}} = \SI{10}{h}$).
As expected, the amount of training data impacts the forecast error, which is higher than Figure~\ref{fig:roomeval1}. However, the curve of estimated values is still quite close to the curve of the observed values, thanks to the effective initialization of the model parameters $p^\guess$.
Indeed, through the \textit{guess} field explained in Section~\ref{sec:model}, the user can provide the initial values of the model parameters within the TD of the WT, based on prior contextual knowledge of the use case; such assignments are then automatically loaded by the RDT framework.
Vice versa, Figure~\ref{fig:roomeval3} shows the impact of parameter calibration and training on the performance of the DT. More specifically, we plot with a \textit{yellow} dashed line the estimated temperature values when no learning has been performed at all: in addition, the model parameters are initialized with random $p^\guess$ values, and not changed during the training phase.
We can notice that the expected values diverge significantly from the observed ones.
In the same Figure, the red solid line depicts the case where the training is executed, but  without any initial calibration of the $p^\guess$ parameters.
Comparing Figures~\ref{fig:roomeval1}--\ref{fig:roomeval3}, we can notice that the calibration has a positive impact on the forecast accuracy; however, even without it, the module is able to learn from the incoming data and to improve its performance over time.

Finally, in Table~\ref{tab:mse-room} we summarize the performance for all the configurations considered in Figures~\ref{fig:roomeval1}--\ref{fig:roomeval3}, by reporting the mean squared error (MSE) during the test phase only (i.e., $t> t_{\text{spawn}}$).
The third row (0\% training, Random initial guess) refers to the baseline where no parameter initial calibration and no learning phase have been performed.
Vice versa, it is easy to notice that  both the training time and the tuning of the initial guesses are of paramount importance to minimize the MSE metric (row 2).

\begin{table}
	\centering
	\caption{Mean squared error for experiments plotted in Figure~\ref{fig:roomeval}. The initial guess for each experiment can be either designed by the modelers or randomly assigned.}
	\begin{tabular}{lll}
		\hline
		\% Training & Initial guess & MSE \\ \hline
		20 (\SI{10}{h})         & Designed      & 0.822   \\
		70 (\SI{34}{h})         & Designed      & 0.16   \\
		0  (\SI{0}{h})        & Random        & 17.649   \\
		70 (\SI{34}{h})         & Random        & 6.376   \\ \hline
	\end{tabular}
	\label{tab:mse-room}
\end{table}

\subsection{Real testbed: quadrocopter}
\label{ssec:testuav}

\begin{figure}[t]
	\centering
	\includegraphics[width=0.89\linewidth]{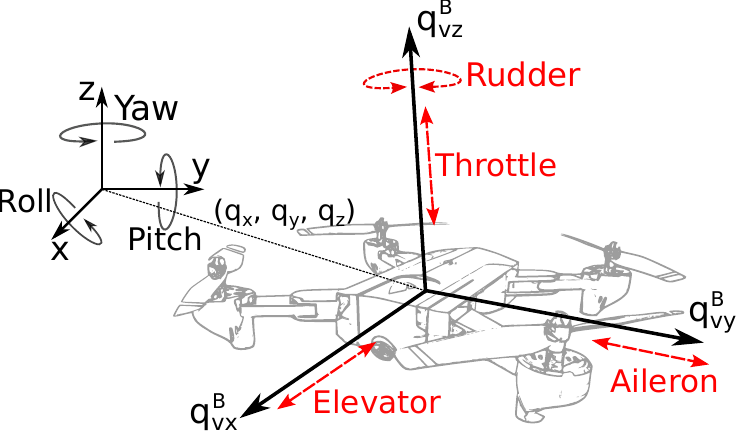}
	\caption{Coordinate system for the quadcopter. The \textit{red} elements represent the joystick inputs.}
	\label{fig:quadcopter}
	\vspace{-3mm}
\end{figure}

Differently from the previous experiment, in this Section, we evaluate the proposed framework in a real testbed scenario. 
The case study is the DT  of a  quadcopter (in our case, a \textit{DJI Mini 2} drone\footnote{\url{https://www.dji.com/mini-2}}), flying in an outdoor environment.
A quadcopter is a small-scale helicopter with four rotors directed upwards and placed in a square formation from the quadcopter's center of mass.
The quadcopter movements can be controlled by setting the angular velocities of the rotors which are spun by electric motors. 
In the literature,  several models have been proposed in order to characterize the quadcopter's kinematics, at different granularity levels.
Interested readers can refer to \cite{zhang2014survey} for a detailed review.
In this work, we used a second-order linear dynamics model in the body reference frame \cite{zhang2014survey}.
Specifically, let $q$ be the target quadcopter.
The latter is defined by its position in global reference frame with the triple $(q_x, q_y, q_z) \in \mathbb{R}^3$ and by its attitude with $(q_{\phi}, q_{\theta}, q_{\psi}) \in \mathbb{R}^3$. Here, $(q_{\phi}, q_{\theta}, q_{\psi})$ are the three Euler angles modeling the \textit{yaw} ($-\pi \leq q_{\psi} \leq \pi$), the \textit{pitch} ($-\nicefrac{\pi}{2} \leq q_{\theta} \leq \nicefrac{\pi}{2}$), and the \textit{roll} ($-\nicefrac{\pi}{2} \leq q_{\phi} \leq \nicefrac{\pi}{2}$), which represent the orientation of the quadcopter. 
In the experiments, latitude and longitude are remapped to local $q_x$, $q_y$ coordinates.
The velocity of $q$ in global reference frame is defined by $q_v = (q_{vx}, q_{vy}, q_{vz}) \in \mathbb{R}^3$ and in the body frame reference by $q^B_v = (q^{B}_{vx}, q^{B}_{vy}, q^{B}_{vz}) \in \mathbb{R}^3$ (see Figure~\ref{fig:quadcopter} for a visual description of the variables).
Here, we assume a small angle approximation that yields to $q_{\phi} = q_{\theta} = 0$.
The angular velocity for the yaw is defined by $q_{v\psi}$.
Finally, the following rotation matrix $\mathrm{RT}(q_{\psi})$ is used to  convert the copter velocity from the body frame reference to the inertial frame one:
\begin{equation*}
	\begin{bmatrix}
		q_{vx} \\ q_{vy}
	\end{bmatrix}
	= \mathrm{RT}(q_{\psi}) \cdot 
	\begin{bmatrix}
		q^B_{vx} \\ q^B_{vy}
	\end{bmatrix}
	=\begin{bmatrix}
		\cos q_{\psi} & -\sin q_{\psi} \\ \sin q_{\psi} & \cos q_{\psi}
	\end{bmatrix}
	\cdot 
	\begin{bmatrix}
		q^B_{vx} \\ q^B_{vy}
	\end{bmatrix}
\end{equation*}
The model described so far defines the actual state of the quadcopter $q$, i.e., its position, altitude, and velocity.
The  quadcopter is controlled by a remote controller whose modeled inputs are: 
\begin{itemize}
	\item \textit{throttle} \textit{Th}, that commands climb/decent movement of the copter. Positive values of \textit{Th} are for climbing request, descent otherwise.
	\item \textit{rudder} \textit{Ru}, that commands the yaw movements. Positive values of \textit{Ru} are for the clockwise rotation of the copter.
	\item \textit{elevator} \textit{El}, that commands the pitch, i.e., the forward and backward movements. Here $El > 0$ means forward, backward otherwise. 
	\item \textit{aileron} \textit{Ai}, that commands the roll, i.e., the left and right movements, where $Ai > 0$ means left movements, right direction otherwise. 
\end{itemize}
The aforementioned user inputs constitute the set of writable variables $W$ of the device.
The full quadcopter model description and its translation in TD is described in Section~\ref{sec:appendixuav}.

Using the resulting TD,
we deployed the WT of the DJI quadcopter and connected it to the proposed RDT framework.
During the flight, we collected the drone's data thanks to the DJI APIs that allow downloading the raw data of the onboard sensors. More specifically, we measured the following quantities: \textit{latitude}, \textit{longitude}, and \textit{altitude} for the drone position (that are remapped to local $(x,y,z)$ coordinates); \textit{compass} for the drone's attitude (where \textit{compass} defines the \textit{yaw}), and \textit{elevator}, \textit{aileron}, \textit{throttle}, \textit{rudder} as the joystick user's inputs.
%\\
The experiments  aim to validate the capability of the RDT framework related to the following aspects: (\textit{i}) the overall ability of the proposed TD extension to model complex IoT systems like the proposed one,
(\textit{ii}) the accuracy of the learning phase and the impact of different parameters such as the training length; (\textit{iii}) the effectiveness of the DT to support what-if analysis and hence to avoid harmful operations, thanks to the capability of simulating the system reactions to user's commands.

After the take-off of the quadcopter, the RDT framework starts acquiring raw data from the drone and activates the \textit{model training} phase. 
The take-off and the landing procedures are considered special movements hence they are not taken into account during the training phase.
We tested  two DTWT spawning events at a distance of \SI{10}{s} one from the other in order to evaluate the impact of the training length.

\begin{figure*}%
	\centering
	\subfloat[]{\includegraphics[width=0.33\textwidth]{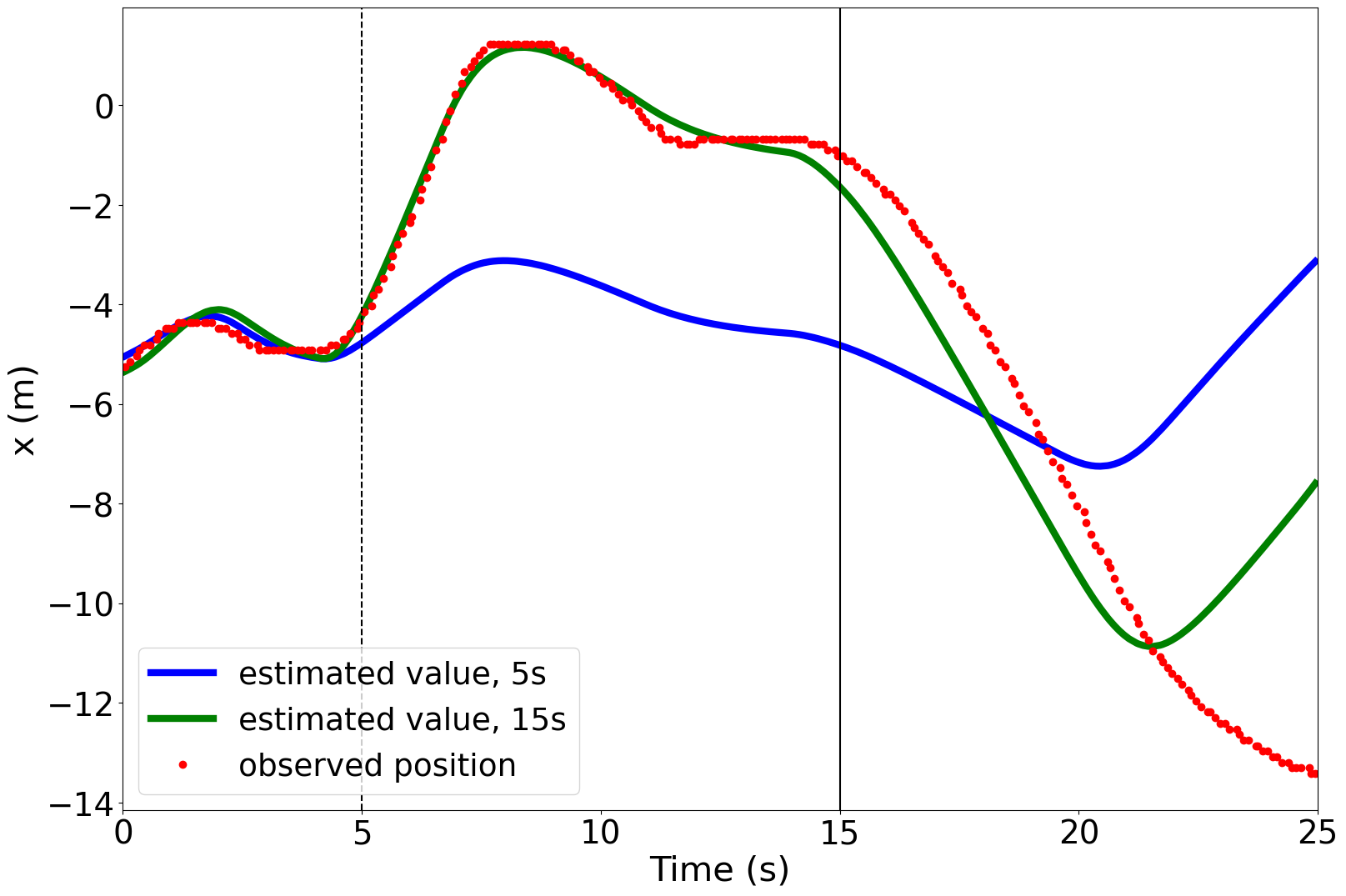} \label{fig:drone-exp-t1}}
	\subfloat[]{\includegraphics[width=0.32\textwidth]{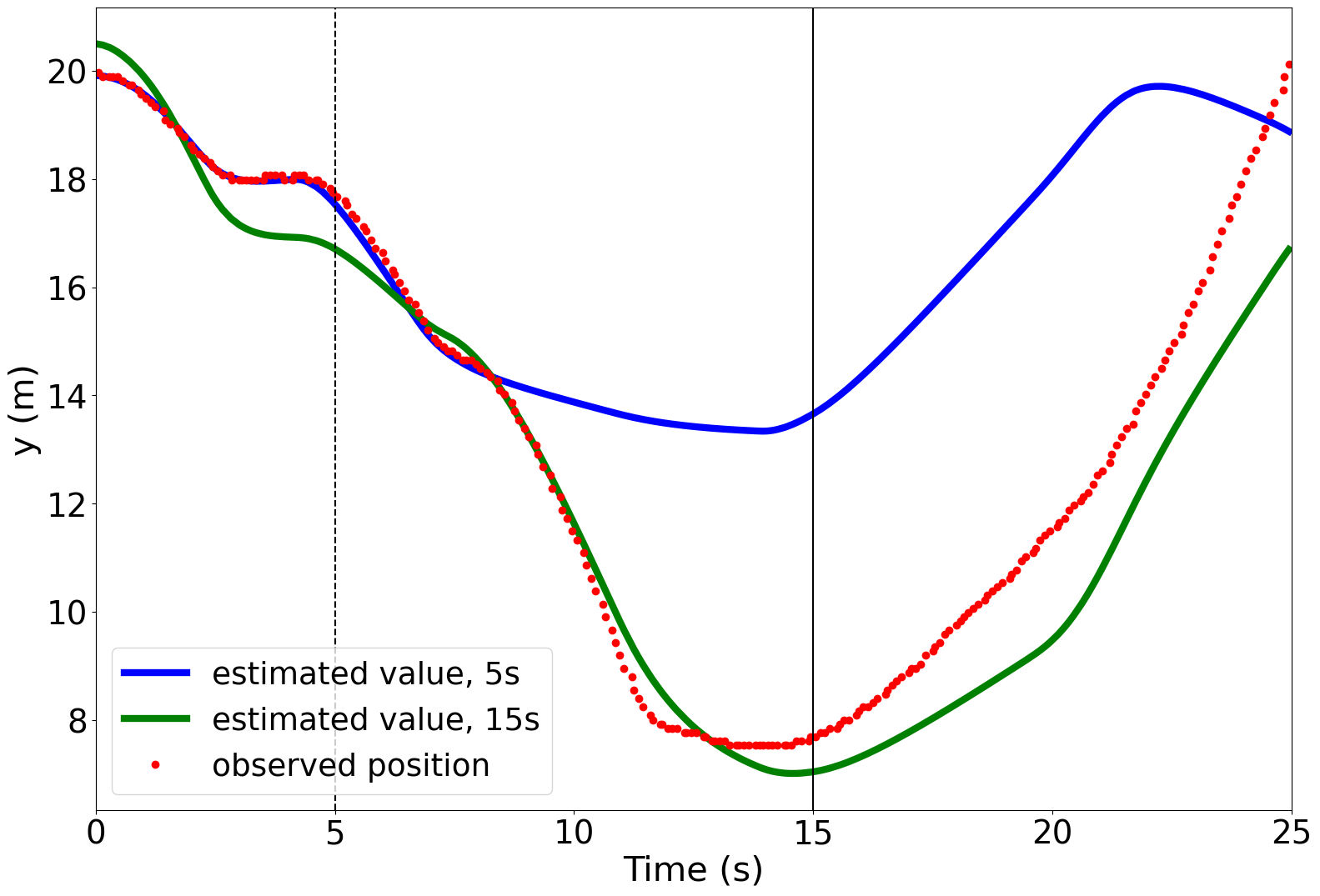} \label{fig:drone-exp-t2}}
	\subfloat[]{\includegraphics[width=0.33\textwidth]{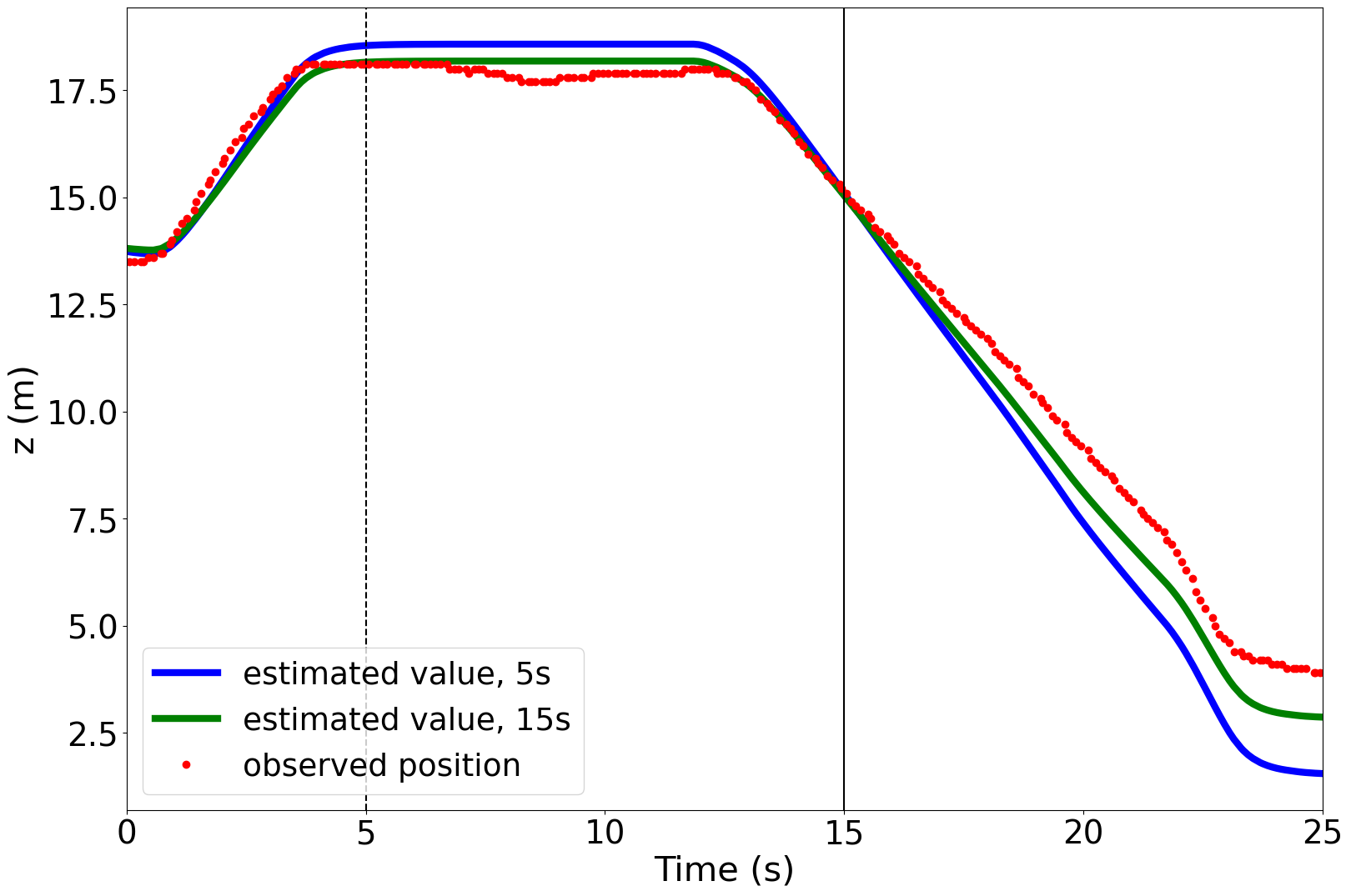} \label{fig:drone-exp-t3}}
	\caption{The predicted values of $x$, $y$, and $z$ from the spawned DTWTs are shown in Figures~\ref{fig:drone-exp-t1}, \ref{fig:drone-exp-t2}, and \ref{fig:drone-exp-t3}, respectively. In the Figures, the \textit{red} dots show the collected data from the real drone, the \textit{blue} lines are the predicted values from the DTWT spawned after \SI{5}{s} (dashed vertical line), and the \textit{green} lines are the predicted values from the DTWT spawned after \SI{15}{s} (solid vertical line).}
	\label{fig:droneeval}
\end{figure*}
In Figures~\ref{fig:drone-exp-t1}, \ref{fig:drone-exp-t2}, \ref{fig:drone-exp-t3} we plot the real $(q_x, q_y, q_z)$ values (got from the onboard sensors) with a dashed red line. In the same Figures, we plot the estimated position
when the DTWTs are spawned at $t_{\text{spawn}}$= \SI{5}{s} (blue solid line) and at   $t_{\text{spawn}}$=\SI{15}{s} (green solid line). It is easy to notice that the three curves almost overlap during the training phase, i.e., when $t<$ $t_{\text{spawn}}$. Vice versa, the curves of observed values and of predicted values with $t_{\text{spawn}}$ = \SI{5}{s} diverge significantly during the test phase when the DT is used as a standalone component: this is a clear symptom that the training phase was too short. Indeed, with $t_{\text{spawn}}$ = \SI{15}{s}, the DT is able to forecast the trajectory of the quadcopter in a quite accurate way for all three axes.

\begin{figure}[t]
	\centering
	\includegraphics[width=0.98\linewidth]{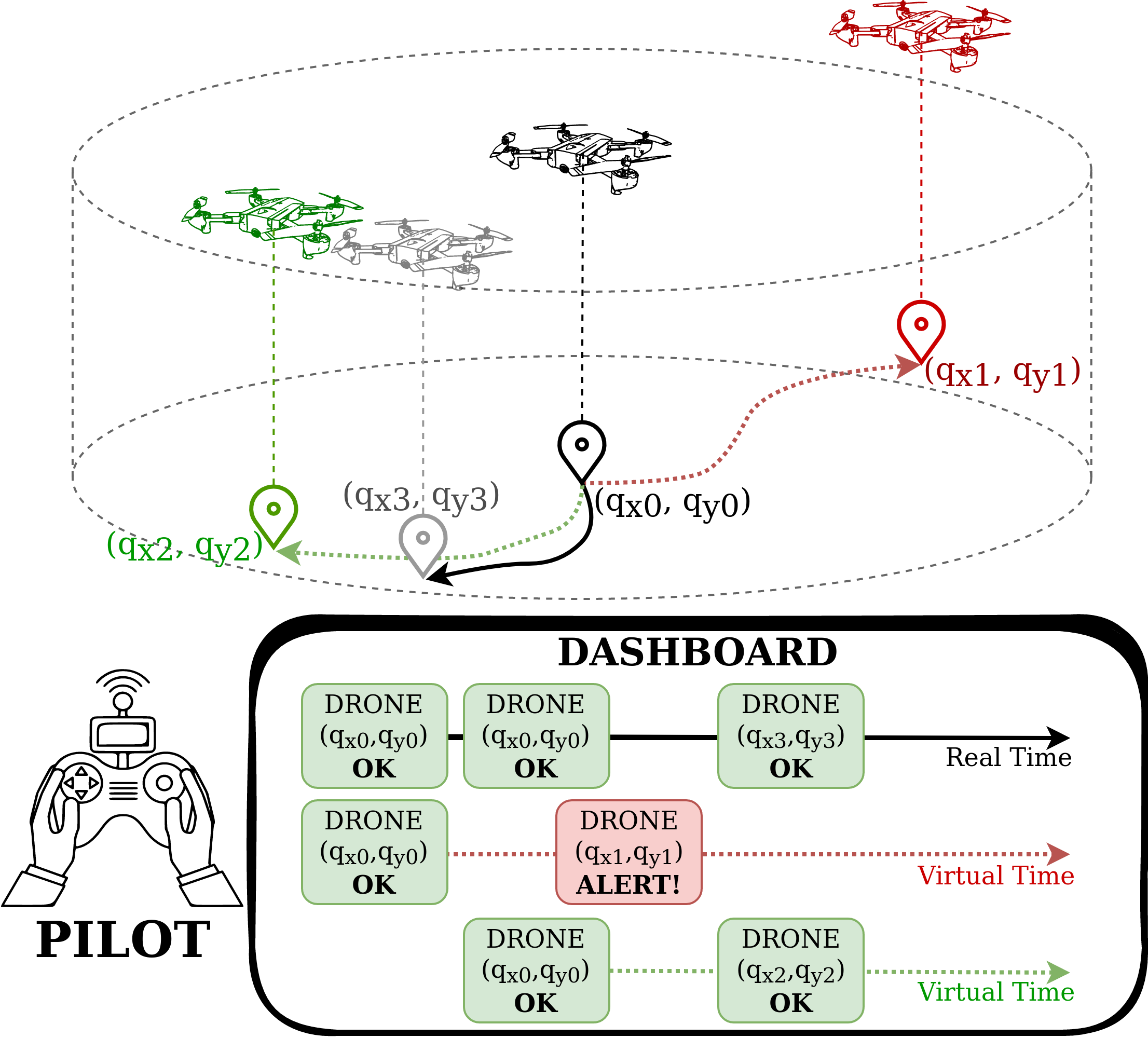}
	\caption{A use-case scenario for the RDT framework. The pilot checks the control commands on the DTWT before issuing them to the drone.}
	\label{fig:scenario-geo}
\end{figure}

\begin{figure*}%
	\centering
	\subfloat[]{\includegraphics[width=0.33\textwidth]{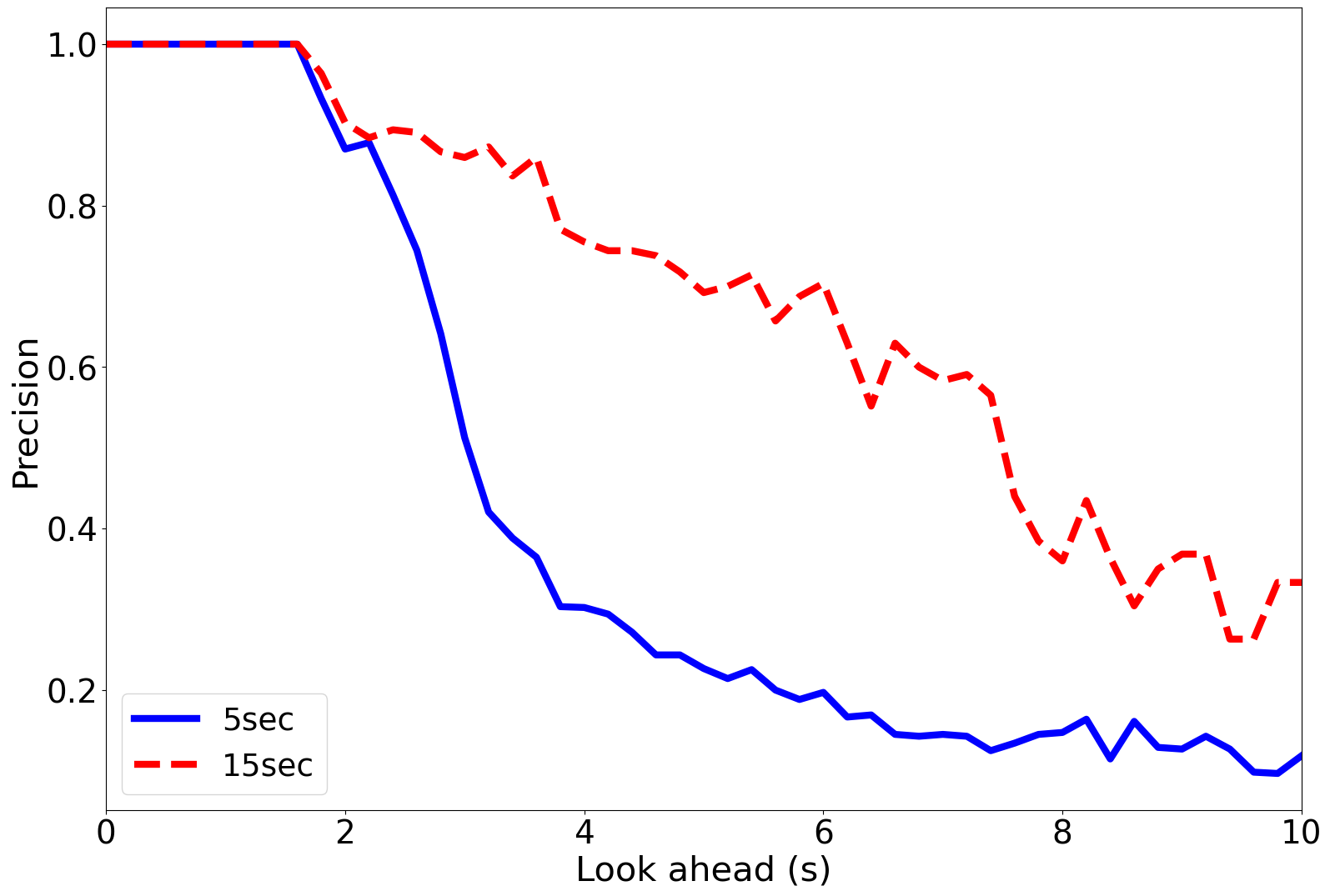} \label{fig:drone-exp-t5}}
	\subfloat[]{\includegraphics[width=0.33\textwidth]{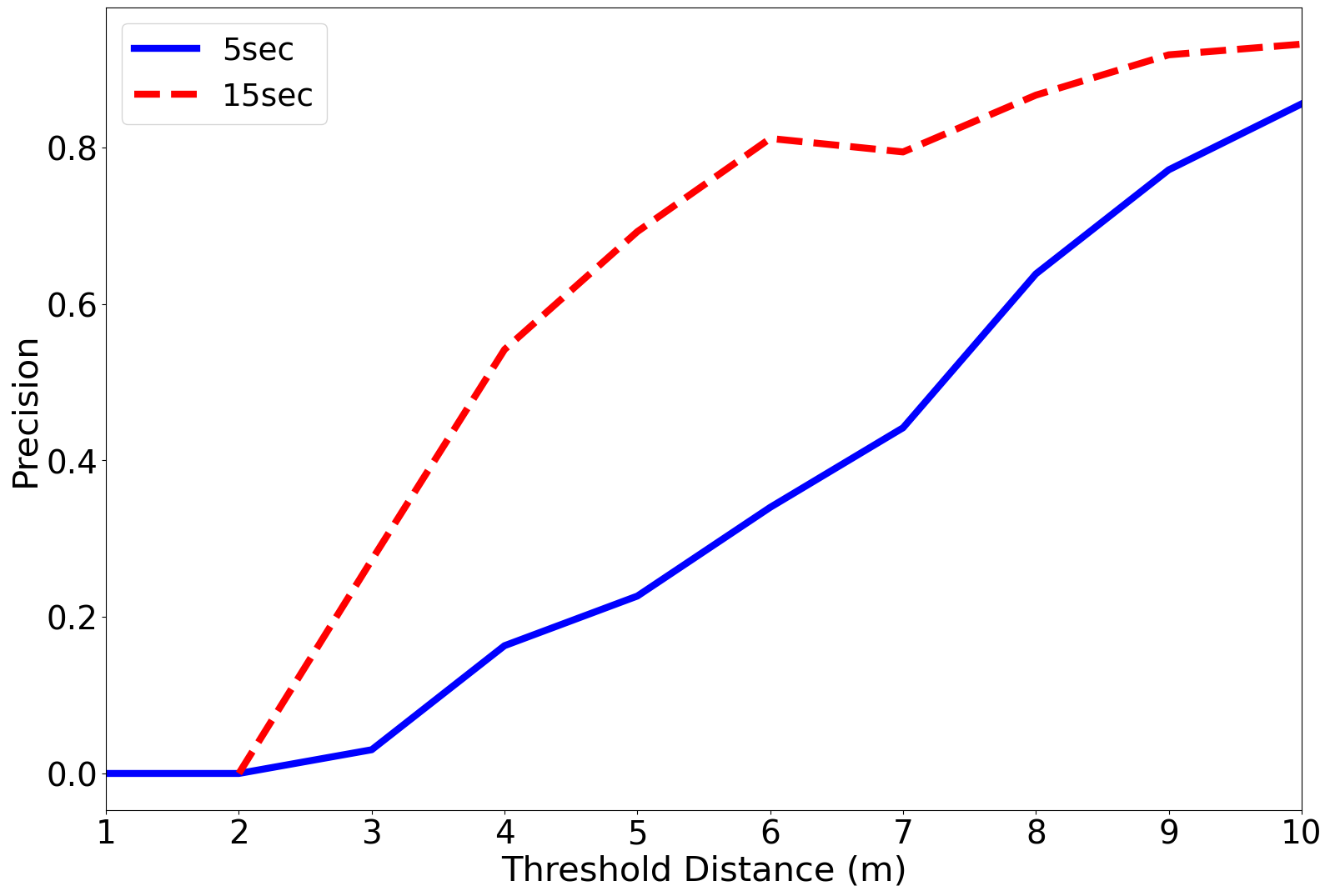} \label{fig:drone-exp-t6}}
	\subfloat[]{\includegraphics[width=0.33\textwidth]{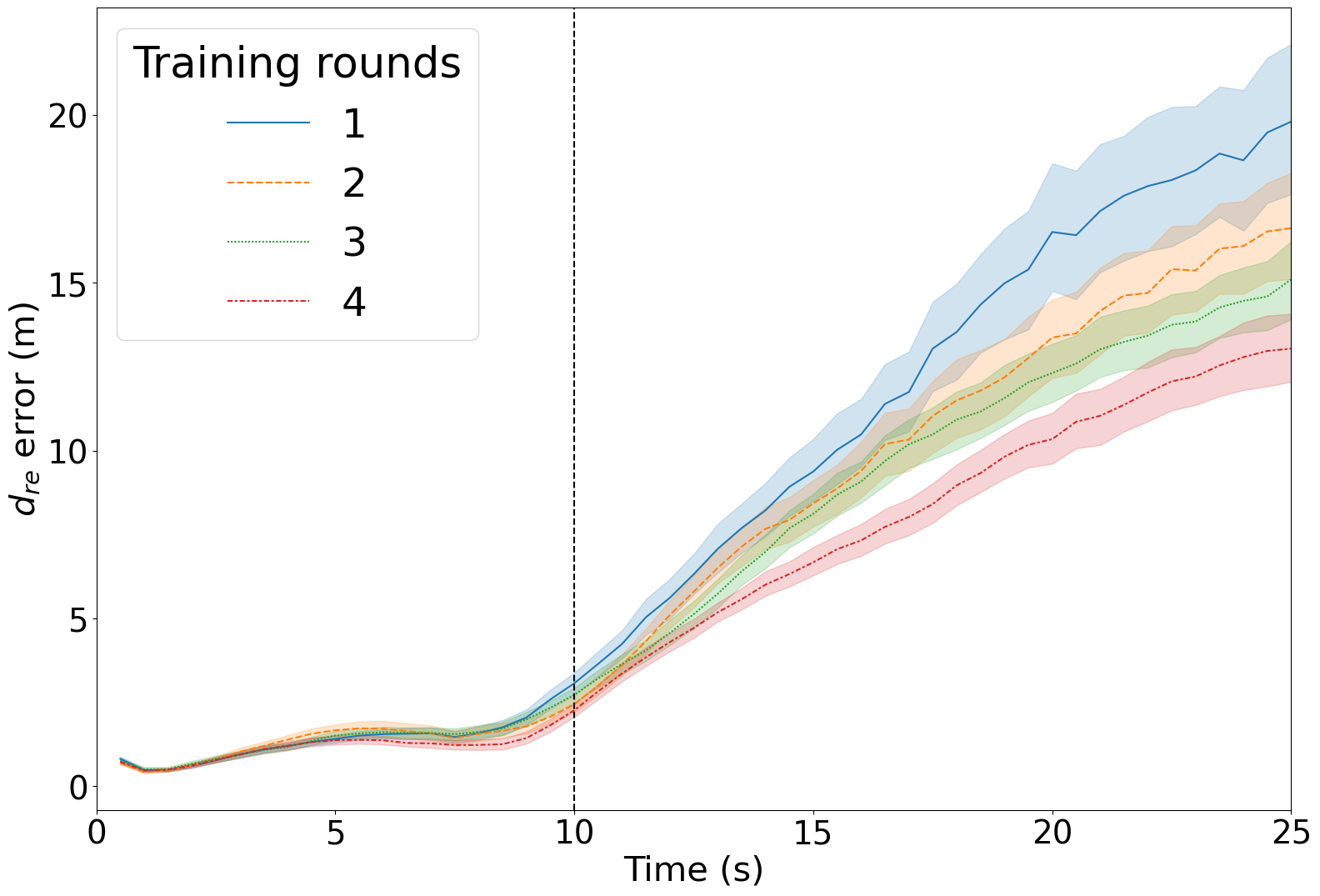} \label{fig:drone-exp-t7}}
	\caption{The \textit{precision} index is depicted in Figures~\ref{fig:drone-exp-t5} and \ref{fig:drone-exp-t6} by varying the look-ahead time ($t_{\mathrm{la}}$) and the distance threshold ($d_{\mathrm{thr}}$), respectively. Figure~\ref{fig:drone-exp-t7} shows the distance error $d_{\mathrm{re}}$ over time for consecutive training rounds.}
	\label{fig:droneeval-error}
\end{figure*}

In the next experiment, we investigate the possibility to deploy the DT in parallel with the actual IoT system and to use it for real-time what-if analysis.
More in detail, in Figure~\ref{fig:scenario-geo} we envision the usage of the RDT framework by a drone pilot to safely control his/her device. We consider a scenario with geofence areas, i.e., flight zones from which the drone is not allowed to leave.
After the definition of such areas, the user generates a DTWT of the drone through our RDT framework, and he/she tests any pilot command on the DTWT before executing them on the real drone.
In Figure~\ref{fig:scenario-geo}, the drone is in position $(q_{x0}, q_{y0})$; the user provides a sequence of throttle/rudder/elevator/aileron commands, with the relative time intervals, to the DTWT.
The latter returns an \textit{alert} warning since, based on the  positions estimated by the trained behavioral model, the drone would be located in position $(q_{x1}, q_{y1})$ outside the geofence area.
As a result, the user inputs another sequence of commands which are validated by the DTWT (returning position $(q_{x2}, q_{y2})$).
Finally, he/she issues them to the real drone which moves to the position $(q_{x3}, q_{y3})$, remaining inside the geofence area.

Let $d_{\mathrm{thr}}$ be the radius of a circular geofence area centered on the actual position of the drone. Given a set of control commands, the DT must estimate the future position of the drone, and, in addition, detect the events in which the drone will move outside the geofence area. Let $t_{\mathrm{la}}$ be the look-ahead time that specifies the time offset in the future when the drone position must be predicted. As the evaluation metric for this experiment, we used the \textit{precision} index. The latter is the ratio of correctly predicted positive observations (true positive) to the total predicted positive observations (true positive plus false positive), where high precision relates to a low-false positive rate. In this case, such a ratio is critical because it expresses the times in which the system failed to detect crossing events of the geofence borders. More formally the index is calculated as
\begin{equation*}
	\text{precision} = \frac{\mathrm{TP}}{\mathrm{TP}+\mathrm{FP}}
\end{equation*}
where TP (true positive) represents the true positive values, i.e., when the DTWT correctly predicts that the drone will remain inside the geofence area after $t_{\mathrm{la}}$ seconds. The value of FP (false positive) indicates the number of cases when the  DTWT predicts that the drone will stay inside the area while the real position is outside. 

In Figures~\ref{fig:drone-exp-t5} and \ref{fig:drone-exp-t6}, we plot the \textit{precision} index when varying the look-ahead time $t_{\mathrm{la}}$ and the geofence area $d_{\mathrm{thr}}$, respectively.
Figure~\ref{fig:drone-exp-t5} shows the index for two different training time, $t_{\text{spawn}} = \SI{5}{s}$ and $t_{\text{spawn}} = \SI{15}{s}$, where $d_{\mathrm{thr}} = \SI{5}{m}$.
It is easy to notice that, like for the previous experiments, the training length impacts significantly the estimation of future values. Moreover, the higher the $t_{\mathrm{la}}$ look-ahead time, the lower the prediction index. This is due to the accumulated errors  as the DTWT moves to future time intervals.
However, when the DT is employed with $t_{\text{spawn}} = \SI{15}{s}$, the number of harmful events ---in our case geofence exit--- can be considerably reduced compared to a case when the DT is not employed at all. 
The same trends can be observed in Figure~\ref{fig:drone-exp-t6} where $t_{la} = \SI{5}{m}$.
The radius of the geofence area impacts the precision value in a way that false positive values are less frequent when the target area is larger.

The last experiment shows the possibility for the RDT framework to continuously improve the accuracy of the DTWT model over time. As described in the previous Sections, the behavioral model for the DT is based on a set of free variables $\alpha^D_\star$ that must be properly tuned by the RDT framework.
In the previous experiments, each training stage was initialized with the $P^0$ that are read from the WT TD. On the contrary, in this experiment, we used the same estimated parameters set during consecutive training phases. More formally, let $\widehat{P}^1, \widehat{P}^2, \dots$ be the trained parameters set after $1, 2, \dots$ training rounds that used $P^0_1, P^0_2, \dots$ as initial parameters guess, respectively. To exploit previous knowledge, in each training round we used as initial guess the trained parameters of the previous round, i.e., $P^0_i = \widehat{P}^{i-1}$, $\forall i > 1$ with $P^0_1 = P^0$. 
The rationale of this experiment is to evaluate the capability of the RDT framework to continuously train its DT model during its lifetime.
In Figure~\ref{fig:drone-exp-t7} we show the distance error between the real drone and the predicted position ($d_{\mathrm{re}}$) after consecutive training rounds. Here, the training time is set to \SI{10}{s}. We can notice that the error decreases significantly over the training rounds, hence justifying the effectiveness of continuous learning mechanisms.
We plan to further elaborate on this feature of the RDT framework as future works.

\section{Conclusions and Future Works} \label{sec:conclusion}
In this paper we presented the Relativistic Digital Twin (RDT) framework, which is able to replicate a physical asset, being it a single system or a System of Systems (SoS) into an abstract Digital Twin Web Thing (DTWT) that emulates its behavior and its reaction to events.
The resulting framework has the capability of (\textit{i}) generating heterogeneous and generic Digital Twins that are not necessarily tied to the use case on which they operate, and (\textit{ii}) interoperating with well-known IoT frameworks that use established standards.
To this aim, we targeted an initial modeling of a real-world system based on the Web of Things (WoT) standard promoted by the W3C. As a research contribution, we  proposed to extend the Thing Description (TD) component of the WoT standard in order to take into account the system behavior as a set of  algebraic expressions or differential equations. We presented the new TD vocabulary and all the components of the RDT framework allowing to generate a trained behavioral model by fitting the model parameters to the observed data. The RDT framework was evaluated in two different scenarios, namely the DT of a simulated smart home environment with heterogeneous sensors/actuators and, and the DT of a real-world quadcopter predicting its kinematics over time. In both cases, we assessed the effectiveness of the parameter training process and the capability of the RDT framework to support predictive and what-if-analysis. The work presented in this paper is pioneering and we believe it can trigger the interest of the scientific community in pursuing further extensions of the RDT framework. One of them could be making the generation of these Digital Twins federated so that a SoS of Digital Twins can be composed by building a mashup of pre-generated Digital Twins of appliances that have not yet worked together in the real world. In addition, we plan to include support for additional ML techniques (also model-free) and extend the evaluation to other case studies.

\appendix
\section{Code Snippets - Rooms Scenario}
\label{sec:appendix}
In this Appendix we present three code snippets that exemplify 
the temperature-based use case\footnote{The full TD can be found at \url{https://github.com/UniBO-PRISMLab/relativistic-wot/blob/main/optimizer/thing.jsonld}} presented in Section~\ref{sec:scenario}.
The code below shows the three newly introduced JSON fields for the \texttt{rProp} that represents the temperature in one of the rooms ($T_A$).
As the property is connected with a sensor value, we set \texttt{"valueFrom": "readProperty"}. The \texttt{model} field is specified on top of three local parameters and global parameters 2 and 3. The model also depends on three \texttt{modelInputs}: heaterPower, coolerPower and temp1. temp1 is the temperature of the second room, which is imported into the model through \texttt{input(temp1)}. Since the model of temp1 only specifies ``self'', then we can assume that the optimizer will import the model of the property \texttt{temperature1} as it is. Furthermore, both coolerPower and heaterPower are summed up and imported together through \texttt{sum(inputType(@heatPower))}. This happens because they are responsible (one opposed to the other) for the same phenomenon, i.e., a power causing a temperature variation over time.
Both the models of ``heater'' and ``cooler'' are not imported as they are: the first one is multiplied by a parameter (\texttt{params[0]}, which is different from the \texttt{params[0]} in the model of the temperature) and the second one is multiplied by -1, because its power contributes negatively to the temperature rise.  

\begin{lstlisting}[caption={DTWT TD, temperature}, label=lst:td-temp, language=Json]
...
"valueFrom": "readProperty",
"modelInput": [
{
	"title": "heaterPower",
	"propertyName": "heater",
	"type": "number",
	"model": "params[0] * self | params[0] >= 0.0 | params[0] = 1.0",
	"modelType": "@heatPower"
},
{
	"title": "coolerPower",
	"propertyName": "cooler",
	"type": "number",
	"model": "-self",
	"modelType": "@heatPower"
},
{
	"title": "temp1",
	"propertyName": "temperature1",
	"type": "number",
	"model": "self"
}
],
"model": "dot(self) = params[0] * (params[1] * (global[2] - self) + sum(inputType(@heatPower)) + global[3] * (input(temp1) - self)) | params[0] >= 0.0, params[1] >= 0.0, global[3] >= 0.0 | params[0] = 0.001, params[1] = 0.1, global[2] = 15.0, global[3] = 0.1",
...
\end{lstlisting}

The reason behind how we are treating differently heaterPower and coolerPower lies on top of how they are defined. Below we can see how heaterPower is defined: its model only outputs the actual value of the \texttt{rProp}. This identifies a set of cases where the manufacturer only provides the behavior of the component independently from the environment in which it acts. In fact, in this case, the impact of the heaterPower has to be tuned via a parameter within the modelInputs in Listing~\ref{lst:td-temp}.

\begin{lstlisting}[caption={DTWT TD, heaterPower}, label=lst:td-heater, language=Json]
...
"valueFrom": "readProperty",
"model": "self = value()",
...
\end{lstlisting}

Differently, the coolerPower, as it is defined in Listing~\ref{lst:td-cooler}, indicates a \texttt{rProp} that encloses pretty much all the information about the cooler. In fact, the \texttt{value()} of the cooler is never used, instead the \texttt{rProp} always outputs the value calculated by a model that defines how the cooler would impact an environment on top of two model parameters. Furthermore, the model imports the model of the \texttt{rProp} called ``coolerSetPoint'' as a modelInput. The latter is another \texttt{rProp} that only outputs the power lever of the cooler knob as a number constrained between 0 and 9. 

\begin{lstlisting}[caption={DTWT TD, coolerPower}, label=lst:td-cooler, language=Json]
...
"valueFrom": "model",
"modelInput": [
{
	"title": "coolerSetpoint",
	"type": "number",
	"model": "max(0, min(round(value()), 9))"
}
],
"model": "dot(self) = params[0] * (params[1] * input(coolerSetpoint) - self) | params[0] >= 0.0, params[1] >= 0.0 | params[1] = 0.1, params[0] = 0.1",
...
\end{lstlisting}

\section{Quadcopter Model}
\label{sec:appendixuav}
In this Appendix, we describe in detail the quadcopter model and its relative translation in the TD.

Following the drone's model description of Section~\ref{ssec:testuav} with states and inputs, we define the system dynamics as follows:
\begin{align}
	\dot{q}_x &= q_{vx} \label{eq:dynamic-drone-px}\\
	\dot{q}_y &= q_{vy} \\
	\dot{q}_z &= q_{vz} \\
	\dot{q}_{\psi} &= q_{v\psi} \\
	\dot{q}_{vx}^B &= \alpha^D_1 (\alpha^D_2 El - q_{vx}^B) \\
	\dot{q}_{vy}^B &= \alpha^D_3 (\alpha^D_4 Ai - q_{vy}^B) \\
	\dot{q}_{vz} &= \alpha^D_5 (\alpha^D_6 Th - q_{vz}) \\
	\dot{q}_{v\psi} &= \alpha^D_7 (\alpha^D_8 Ru - q_{v\psi})
\end{align}
complemented by the kinematic relationships:
\begin{align}
	\begin{bmatrix}
		q_{vx}^B \\ q_{vy}^B
	\end{bmatrix} &= \mathrm{RT}(q_{\psi})^\top \begin{bmatrix}
		q_{vx} \\ q_{vy}
	\end{bmatrix} \\
	\begin{bmatrix}
		\dot{q}_{vx} \\ \dot{q}_{vy}
	\end{bmatrix} &= \mathrm{RT}(q_{\psi}) \begin{bmatrix}
		\dot{q}_{vx}^B \\ \dot{q}_{vy}^B
	\end{bmatrix} + \mathrm{RT}^\prime(q_{\psi}) q_{v\psi} \begin{bmatrix}
		q_{vx}^B \\ q_{vy}^B
	\end{bmatrix}  . \label{eq:dynamic-drone-pvx}
\end{align}
Notice that the dynamics are expressed in the drone's body reference frame and that the first-order models for the velocities stem from Newton's basic laws of classical mechanics and the inertial properties of the copter, namely from its mass distribution.
Furthermore, although uncoupled to the body reference frame, the dynamics equations involve different axes of the inertial reference frame.
The model above contains $8$ free variables, $\alpha^D_1, \dots, \alpha^D_8$, which represent the set of parameters $P$ specific to the quadcopter in use, and which must be properly tuned.
Given such a proposed behavioral model, we defined the TD of the WT associated with the quadcopter; the latter will  be used by the RDT framework to generate the DT of the appliance. Due to the length of the resulting TD, we report here only the most significant fragments\footnote{The full TD can be found at \url{https://github.com/UniBO-PRISMLab/relativistic-wot/blob/main/optimizer/thingDrone.jsonld}}.

\begin{lstlisting}[caption={TD of the quadcopter, x-position property}, label=lst:td-drone1, language=Json]
"positionX": {
	...
	"valueFrom": "readProperty",
	"modelInput": [
	{
		"title": "vX",
		"propertyName": "velocityX",
		"type": "number",
		"model": "self"
	}
	],
	"model": "dot(self) = input(vX)"
},
\end{lstlisting}

The snippet  in Listing~\ref{lst:td-drone1} shows the \textit{x-position} property of the drone, i.e., $q_x$.
Similar to $q_y$ and $q_z$, the position of the drone is read directly from the onboard sensors and hence the \texttt{valueFrom} field is set to \texttt{readProperty}.
Following \eqref{eq:dynamic-drone-px} describing the drone dynamics for the \textit{x-position}, we included the \texttt{model} field that reflects the position's dynamic using the \texttt{dot(self)} notation.
We can also notice from Listing~\ref{lst:td-drone1} the usage of the \texttt{modelInput} field that lists the  properties to be used in the \texttt{model}.

\begin{lstlisting}[caption={DTWT TD, x-velocity property}, label=lst:td-drone2, language=Json]
"velocityX": {
	...
	"valueFrom": "model",
	"modelInput": [
	{
		"title": "yaw",
		"propertyName": "yaw",
		"type": "number",
		"model": "self"
	},
	{
		"title": "yawrate",
		"propertyName": "yawrate",
		"type": "number",
		"model": "self"
	},
	{
		"title": "vbX",
		"propertyName": "velocitybodyX",
		"type": "number",
		"model": "self"
	},
	{
		"title": "vbY",
		"propertyName": "velocitybodyY",
		"type": "number",
		"model": "self"
	},
	{
		"title": "abX",
		"propertyName": "accelerationbodyX",
		"type": "number",
		"model": "self"
	},
	{
		"title": "abY",
		"propertyName": "accelerationbodyY",
		"type": "number",
		"model": "self"
	}
	],
	"model": "dot(self) = math.cos(input(yaw)) * input(abX) - math.sin(input(yaw)) * input(abY) - input(yawrate) * (math.sin(input(yaw)) * input(vbX) + math.cos(input(yaw)) * input(vbY))"
},
\end{lstlisting}
Another fragment of the TD is reported in Listing~\ref{lst:td-drone2}, focusing on the modeling of the quadcopter's velocity. More specifically, we show the \texttt{velocityX} property that corresponds to the $q_{vx}$ state of the quadcopter and whose dynamics are described by \eqref{eq:dynamic-drone-pvx}. In this case, the \texttt{valueFrom} field is set to \texttt{model} which means that the property has no real sensor in the drone and hence the value is derived from the model. The \texttt{model} field encodes \eqref{eq:dynamic-drone-pvx} using the syntax described in Section~\ref{sec:model}. Finally, the \texttt{modelInput} field lists the other properties used inside the model.

\section*{Declaration of competing interest}
The authors declare that they have no known competing financial interests or personal relationships that could have appeared to influence the work reported in this paper.

\section*{CRediT authorship contribution statement}
\noindent
\textbf{Luca Sciullo}: Conceptualization, Methodology, Software, Validation, Formal analysis, Investigation, Writing - Original Draft, Writing - Review \& Editing, Visualization. \\
\textbf{Alberto De~Marchi}: Conceptualization, Methodology, Software, Formal analysis, Writing - Original Draft, Writing - Review \& Editing. \\
\textbf{Angelo Trotta}: Validation, Formal analysis, Investigation, Visualization. \\
\textbf{Federico Montori}: Methodology, Writing - Original Draft. \\
\textbf{Luciano Bononi}: Supervision. \\
\textbf{Marco Di~Felice}: Supervision, Funding acquisition, Project administration, Writing - Review \& Editing.

\bibliographystyle{elsarticle-num}
\bibliography{biblio}

\end{document}